\documentclass[aps,pra,twocolumn,amsmath,amssymb,showpacs,10pt]{revtex4-1}

\usepackage{times}

\usepackage[usenames,dvipsnames]{xcolor}
\newcommand{\Brev}[1]{\textcolor{black}{#1}}

\newcommand{\GGrev}[1]{\textcolor{black}{#1}}

\usepackage[english]{babel}
\usepackage{graphicx}
\usepackage{dcolumn}
\usepackage{bm}
\usepackage{color}
\usepackage{amsmath}
\usepackage{relsize,amsmath,dsfont,mathrsfs,empheq,verbatim,upgreek}
\usepackage{etoolbox}
\usepackage[caption = false]{subfig}
\apptocmd{\thebibliography}{\raggedright}{}{}

\usepackage{hyperref}
\usepackage{graphics}
\usepackage{bm}
\usepackage{color}
\usepackage{amscd}
\usepackage{amsfonts}
\usepackage{amssymb}
\usepackage{graphicx}
\usepackage{tabularx}

\newcommand{\mean}[1]{\langle #1 \rangle}
\providecommand{\bgreek}[1]{\mbox{\boldmath$#1$}}
\newcommand{\vett}[1]{\mathbf{#1}}

\newcommand{\Ham}{\mathcal{H}}

\begin{document}
\newdimen\origiwspc%
  \newdimen\origiwstr%

\title{Energy backflow in strongly coupled non-Markovian continuous-variables systems}

\author{G. Guarnieri$^{1,2}$, J. Nokkala $^{3}$, R. Schmidt $^{3,4,5}$, S. Maniscalco $^{3,4}$, B. Vacchini$^{1,2}$}

\affiliation{$^1$\mbox{Dipartimento di Fisica, Universit{\`a} degli Studi di Milano, Via Celoria 16, 20133 Milan, Italy}\\
$^2$\mbox{Istituto Nazionale di Fisica Nucleare, Sezione di Milano, Via Celoria 16, 20133 Milan, Italy}\\
$^3$\mbox{Turku Centre for Quantum Physics, Department of Physics and Astronomy, University of Turku, FIN-20014 Turku, Finland}\\
$^4$\mbox{Center for Quantum Engineering, Department of Applied Physics, Aalto University School of Science, P.O. Box 11000,
FIN-00076 Aalto, Finland}\\
$^5$\mbox{COMP Center of Excellence, Department of Applied Physics, Aalto University School of Science, P.O. Box 11000,
FIN-00076 Aalto, Finland}}

\begin{abstract} 
  By employing the full counting statistics formalism, we characterize
  the first moment of energy that is exchanged during a generally
  non-Markovian evolution in non-driven continuous variables systems. 
  In particular,
  we focus on the evaluation of the energy flowing back
  from the environment into the open quantum system. We apply these results to
  the quantum Brownian motion, where these
  quantities are calculated both analytically, under the weak coupling
  assumption, and numerically also in the strong coupling regime.
  Finally, we characterize the non-Markovianity of the reduced
  dynamics through a recently introduced witness based on the
  so-called Gaussian interferometric power and we discuss its
  relationship with the energy backflow measure.
\end{abstract}
 
\date{\today}
\maketitle

\section{Introduction}

In recent years, much work has been devoted to understand energy flow and transport properties in the context of open quantum systems \cite{Belzig, Bagrets, Flindt, Esposito, Carrega, SMA2016}. These questions naturally arise since all realistic quantum systems are open and interact with their environments, which in many cases can be modelled as generic heat baths consisting of bosonic modes. Both discrete and continuous variables (CV) open quantum systems have received considerable attention, the topics ranging from controlling heat flow at a microscopic level to better understanding the high efficiency of photosynthesis in biological systems \cite{EngelNAT2007, AspuruGuzikJCHEMPH2011, HanggiRMP2009, AbahPRL2012, Kosloff2013Entropy, Golubev2013PRB, Correa2014SciRep}.
Despite these and other related statistical properties are often described within the so-called Born-Markov approximation \cite{Ren}, many of such systems clearly show memory effects during the dynamics. Moreover, many of the earlier studies still stick to the weak coupling assumption, usually necessary to derive a closed master equation for the statistical operator of the reduced system.
 
In this work, we consider the energy exchange dynamics in non-driven
open CV systems using full counting statistics methods, referring to a
recently introduced measure of energy backflow \cite{OurWork2015},
which quantifies the total amount of energy flowing back to the system
from the heat bath. In particular, we calculate it for a model of quantum Brownian motion (QBM) both in the weak coupling regime, using an analytic approach, and in the strong coupling regime by employing a numerical strategy based on exact diagonalization of a large but finite heat bath.
In this context, we study the role of coupling strength, temperature
and cut-off frequency in the behavior of the energy backflow measure. We also separately consider the dynamics of the energies of system, bath and interaction for different regimes of coupling strength, showing a qualitative change in the dynamics when moving to the strong coupling regime. Finally, we show that, in the considered range of parameters, the reduced dynamics is indeed non-Markovian by using a recently introduced witness which is based on the non-monotonicity of Gaussian interferometric power, which in turn depends on general discord-like correlations between the system and an isolated ancilla.

The paper is organized as follows. In Sec. \ref{sec:Formalism} we recall the two-time measurement protocol formalism and the notion and quantifier of energy backflow, extending it to CV systems. We then apply this method to the QBM in Sec. \ref{sec:QBM}, where we first explicitly characterize analytically all the quantities in the weak-coupling regime and then we evaluate them in the strong coupling regime by means of a numerical approach, showing explicitly the agreement of the two methods in the weak coupling case. In Sec. \ref{sec:Markov} we study the non-Markovianity of the dynamics. Conclusions are finally drawn in Sec. \ref{sec:Conclusions}.

\section{Two-time measurement protocol and Energy Backflow in continuous variable systems}\label{sec:Formalism}

\begin{figure}[htbp!]
\begin{center}
\includegraphics[width=0.9\columnwidth]{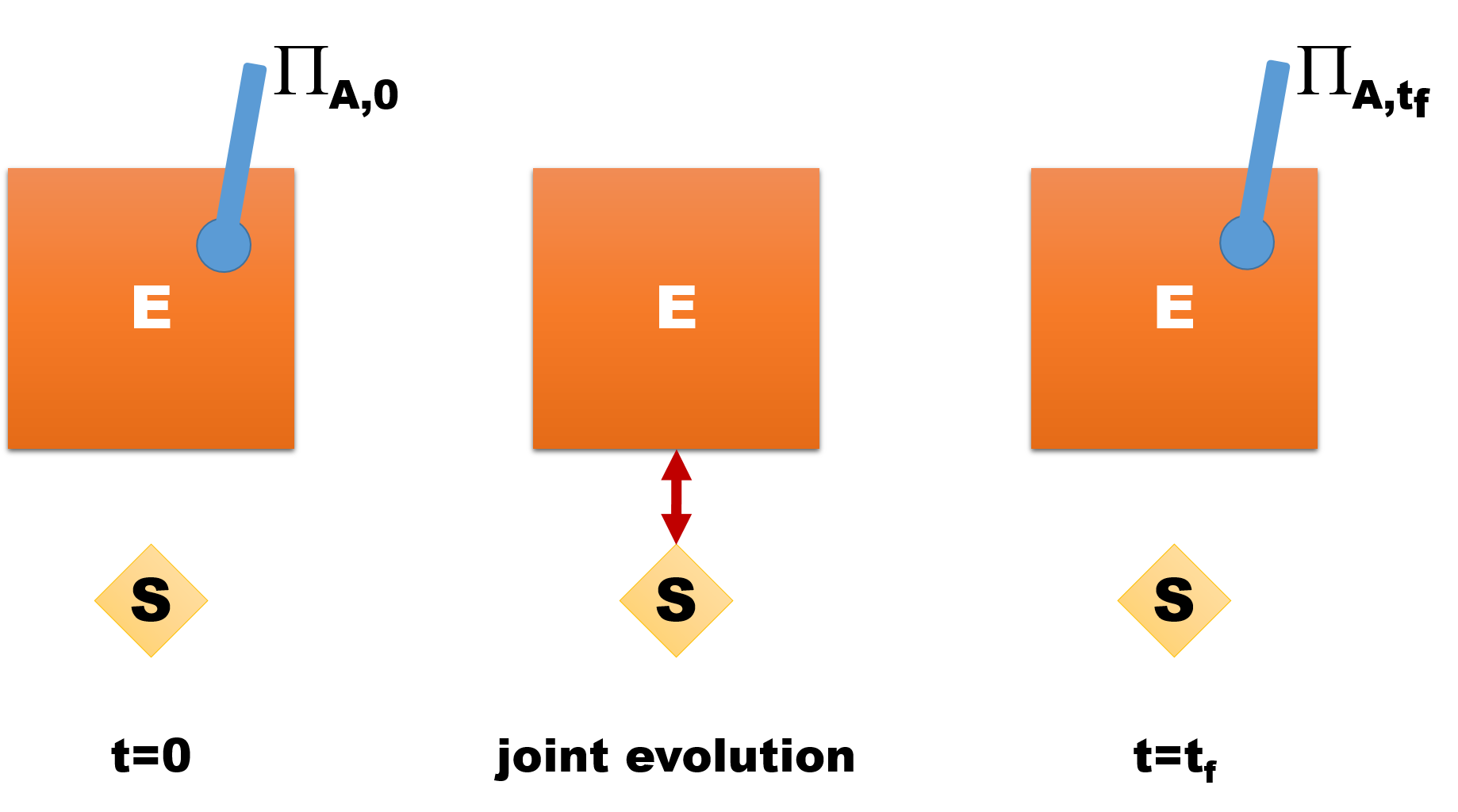}
\caption{(Color online) Consider a system of interest \textit{S} interacting with its environment \textit{E}, to which it is assumed to be decoupled at time $t \leq 0$. At time $t=0$ a generic observable $A$ of the environment is measured through a projective measurement $\Pi$, thus obtaining a certain outcome $a_0$ belonging to its spectrum. \Brev{Immediately after the interaction is switched on and the overall system evolves up to a time $t_f$,} at which time we switch off the interaction and perform another measurement of the observable $A$, this time obtaining another outcome $a_t$. In our case $A$ will be the energy of the environment $\Ham_E$.}
\label{fig:theme}
\end{center}
\end{figure}

To calculate the energy flow between a system of interest \textit{S} interacting with its environment \textit{E}, we employ the  full-counting statistics (FCS) formalism, which has been first introduced in the framework of quantum transport \cite{SS, LL, LLY, ABGK}
and then extended to more general settings, including energy transfers \cite{Belzig, Bagrets, Flindt, Esposito}. This method provides the cumulants of the probability distribution for the change in a generic observable of the environment based on a two-time measurement protocol (see Fig. \ref{fig:theme}).
For open quantum systems, in the CV case, the cumulant-generating function can be written as 
\begin{equation}\label{charchar}
 S_t(\eta)= \ln \mathrm{Tr}_{S} \left[\rho_S(\eta,t)\right] = \ln \chi^{(\eta)}(0,0,t),
\end{equation}
where $\chi$ is the characteristic function associated to the modified density operator $\rho_S(\eta,t)$ \cite{Carmichael,Puri,Olivares}
\begin{equation}\label{charfunc}
\chi \left[\rho_S(\eta,t)\right](\lambda,\lambda^*)\!\equiv\!\chi^{(\eta)}(\lambda,\lambda^*,t)\!=\!\!\mathrm{Tr}_{S} \left[\rho_S(\eta,t) e^{\lambda a^{\dagger}\!-\!\lambda^*\!a}\right],
\end{equation}
with $ a,a^{\dagger}$ denoting the annihilation and creation operators relative to the system.
The modified density operator:
\begin{equation}\label{eq:3}
\rho_S(\eta,t) \equiv \mathrm{Tr}_{E} \left\{U_{\eta/2}(t,0) \rho_{SE}(0) U^{\dagger}_{-\eta/2}(t,0)\right\}
\end{equation}
evolves according to
$
U_{\eta}(t,0) \equiv e^{i\eta \Ham_E} U(t,0) e^{-i\eta \Ham_E},
$
which describes the evolution conditioned to the two time measurements of the environmental energy $\Ham_E$.
When $\eta$ is set to zero, we retrieve both the usual evolution operator $U_{\eta=0}(t,0) = U(t,0) $ and the statistical operator $\rho_S(\eta=0,t) = \rho_S(t)$. However, we emphasize that the system's operator $\rho_S(\eta,t)$ is not a statistical operator (apart from the initial time, when it coincides with $\rho_S(0)$), since its trace is not necessarily normalized to $1$.

Within this framework, the time-dependent first moment of the energy transfer is given by
\begin{equation}\label{enmeanchar}
\langle \Delta q \rangle_t = \frac{\partial \chi^{(\eta)}(0,0,t)}{\partial (i\eta)}|_{\eta=0},
\end{equation}
while the energy flow per unit of time $\theta(t)$ is
\begin{equation}\label{thetachar}
 \theta(t) = \frac{\partial \dot{\chi}^{(\eta)}(0,0,t)}{\partial (i\eta)}|_{\eta=0},
\end{equation}
where the dot denotes the time derivative.
$\theta(t)$ provides the rate by which the system and its environment exchange energy and, more specifically, $\theta(t) > (<) 0$ indicates an increment (decrement) in the environmental energy, i.e. an energy flow from the reduced system (environment) to the environment (reduced system).
In the Born-Markov semigroup limiting case, $\theta(t)$ becomes a monotonic function of time, thus indicating a steady energy flow from the higher to the lower temperature system \cite{Ren,OurWork2015} that vanishes in the case system and environment start with the same initial temperature. 
Beyond the Born-Markov description, the energy flow becomes an oscillating function of time, whose behavior can strongly vary depending on the various parameters characterizing the dynamics.
\GGrev{In particular, we speak of regions of \textit{energy backflow from the environment to the system} whenever, considering situations which in the Born-Markov semigroup approximation would lead to a non-negative steady energy transfer from system to environment, we have that at some time $t$}
\begin{equation}\label{condenb}
\theta(t) < 0.
\end{equation}
Building on this condition, a measure for the total amount of energy
which has flown back from the environment to the system during the
evolution can be introduced as \cite{OurWork2015}
\begin{equation}\label{enbackflow}
\mean{\Delta q}_{back} = \max_{\rho_S(0)}\, \frac{1}{2} \int_0^{+\infty}\,dt\, \left(\left|\theta(t)\right|-\theta(t)\right),
\end{equation}
where the maximization procedure is performed to make it a property of the dynamical map, i.e. independent from the possible choices of initial states of the system.
Note that the integrand of \eqref{enbackflow} is different from zero if and only if $\theta(t)$ is negative and it represents, in principle, a measurable quantity.

\section{Quantum brownian motion}
\label{sec:QBM}

Here we apply the formalism outlined above to the study of the energy transfer in the quantum Brownian motion (QBM), i.e. a quantum harmonic oscillator linearly coupled to an infinite number of bosonic modes. The Hamiltonian of the composite system has the form $\mathcal{H}=\mathcal{H}_S +\mathcal{H}_E+\mathcal{H}_{\rm int}$, with:
\begin{eqnarray}
\label{Ham}
\mathcal{H}_{S}&=&\frac{\omega_0}{2}\left(a^{\dagger} a + 1/2\right),\\ \nonumber
\quad\mathcal{H}_{E}&=&\sum_k\omega_kb^{\dagger}_k b_k , \\ \nonumber
\mathcal{H}_{\rm int}&=& X  \sum_k \left( g_k b^{\dagger}_k + g^*_k b_k \right),
\end{eqnarray}
where $a,a^{\dagger}$ \GGrev{($b_k,b^{\dagger}_k$)} denote the
system's \GGrev{(environmental)} annihilation and creation operators,
$X = 2^{-1/2}(a+a^{\dagger})$, (so that $P = 2^{-1/2} i (a^{\dagger}-a)$), $\omega_k$ is the energy of the $k$th bosonic mode and $g_k$ is the coupling strength between the latter and the system.
In the following, we use natural units, i.e. $\hbar=1$ and $k_{B}=1$.
%


\subsection{Analytical Results in the Weak Coupling Regime}
\label{sec:QBMan}

Under the assumption of weak coupling and secular approximation, a
time-local generalized master equation can be written for the modified
statistical operator of the reduced system $\rho_S(\eta,t)$ according to
\begin{equation}\label{GenME1}
\frac{d}{dt}\rho_S(\eta,t) = \Xi (t) \left[\rho_S(\eta,t)\right] + \mathcal{L}_\eta(t) \left[\rho_S(\eta,t)\right].
\end{equation}
In this expression, the superoperator $\Xi(t)$ has the form
\begin{multline}\label{StdLindblad}
\Xi (t)[\cdot ]\!\equiv\!\!-i\omega_0\left[a^{\dagger}a,\cdot \right ] +\left(\frac{\Delta(t)+\gamma(t)}{2}\right)\!\!\left[2a\cdot a^{\dagger}\! - \lbrace a^{\dagger}a,\cdot\rbrace\right]\\
+\left(\frac{\Delta(t)-\gamma(t)}{2}\right)\left[2a^{\dagger}\cdot a - \lbrace a a^{\dagger},\cdot\rbrace\right],
\end{multline}
and represents the familiar time-dependent Lindblad generator considered in \cite{Intravaia1,Intravaia3}, with
\begin{align}
&\Delta(t) = \frac{1}{2}\int_0^t ds D_1(s) \cos(\omega_0 s),\notag\\
&\gamma(t) = \frac{1}{2}\int_0^t ds D_2(s) \sin(\omega_0 s),\notag\\
&D_1(t) = \Phi(t) + \Phi(-t),\,\, D_2(t) = i\left(\Phi(t) - \Phi(-t)\right),
\end{align}
where
\begin{equation}
\Phi(t) =\!\int_0^{+\infty}\!\!\!\!\!\! d\omega J(\omega)\,[ \mathrm{Coth}\left(\frac{\omega}{2T_E}\right)\cos(\omega t) - i\sin(\omega t)\,]
\end{equation}
is the \textit{environmental correlation function}, $T_E$ is the environmental temperature, and $J(\omega)$ is the
environmental spectral density.
The additional superoperator $ \mathcal{L}_\eta(t) \left[\cdot\right] $ in 
Eq. \eqref{GenME1}, responsible for the non-trace preserving character, has the form
\begin{equation}\label{upsilon}
\mathcal{L}_\eta(t) \left[\cdot\right] \equiv g_+(\eta,t)\, a\,\cdot a^{\dagger} + g_-(\eta,t)\, a^{\dagger}\cdot a,
\end{equation} 
where
\begin{equation}
  \label{eq:1}g_{\pm}(\eta,t)\!=\!\frac{1}{2}\!\int_0^t\!\!ds\! \left[\Delta D_1^{(\eta)}(s) \cos(\omega_0 s) \pm \Delta D_2^{(\eta)}\!(s)\!\sin(\omega_0 s)\right]\notag
\end{equation}
with
\begin{align}
\Delta D^{(\eta)}_{1,2}(t) &\equiv D^{(\eta)}_{1,2}(t) - D_{1,2}(t),\notag\\
D^{(\eta)}_1(t) &= \Phi(t-\eta) + \Phi(-t-\eta),\notag\\
D^{(\eta)}_2(t) &= i\left(\Phi(t-\eta) - \Phi(-t-\eta)\right).
\end{align}
After performing some calculations, detailed in Appendix \ref{app:A}, we are lead to the following expression for the energy flow per unit of time
\begin{multline}\label{thetaFINAL}
\theta(t) = 2\sigma(t)\,\left( \frac{1}{2} D_2(t)\cos(\omega_0 t) + \omega_0 \gamma(t)\right) \\
+  \frac{1}{2} D_1(t)\sin(\omega_0 t) - \omega_0 \Delta(t),
\end{multline}
where $\sigma(t)$ denotes the well-known solution for the covariance matrix of the system \cite{Intravaia1,Intravaia3,Intravaia2,Adesso1}
\begin{equation}\label{sigmasol}
\sigma(0,t) = e^{-2\int_0^t ds\gamma(s)} \left(\sigma(0,0) + \int_0^t\,ds \Delta(s) e^{2\int_0^s d\tau\gamma(\tau)} \right).
\end{equation}
In what follows, we will consider an ohmic spectral density with exponential cut-off $\Omega$
\begin{equation}
J(\omega) = \lambda\omega e^{-\frac{\omega}{\Omega}},
\end{equation}
with $\lambda$ denoting the coupling strength.
As a consequence, the functions $D_1(t)$ and $D_2(t)$, respectively known as \textit{noise} and \textit{dissipation} kernels \cite{Breuer2002}, have the expressions:
\begin{align}\label{eq:D1D2explicit}
&D_1(\tau)\!\!=\!\!2\lambda\!\left[\Omega^2\frac{(\Omega\tau)^2\!-\!1}{(1\!+\!(\Omega\tau)^2)^2}\!+\! 2 T_E^2\!\mathit{Re}\!\left[\psi'\!\! \left(\frac{T_E(1+i\Omega\tau)}{\Omega}\!\right)\!\right]\!\right] \notag\\
&D_2(\tau) =\frac{4\lambda\Omega^3 \tau}{(1+(\Omega\tau)^2)^2},
\end{align}
where ${\psi'(z)} $ is the derivative of the Euler digamma function $\psi(z) = {\Gamma'(z)}/{\Gamma(z)}$.

Note that the initial effective temperature $T_S$ of the system must be chosen to be greater or equal to the initial environmental temperature $T_E$ to ensure the energy backflow is not in the direction of the temperature gradient \cite{OurWork2015}.

\begin{figure}[htbp!]
\begin{center}
{\bf (a)}\\
\includegraphics[width=0.95\columnwidth]{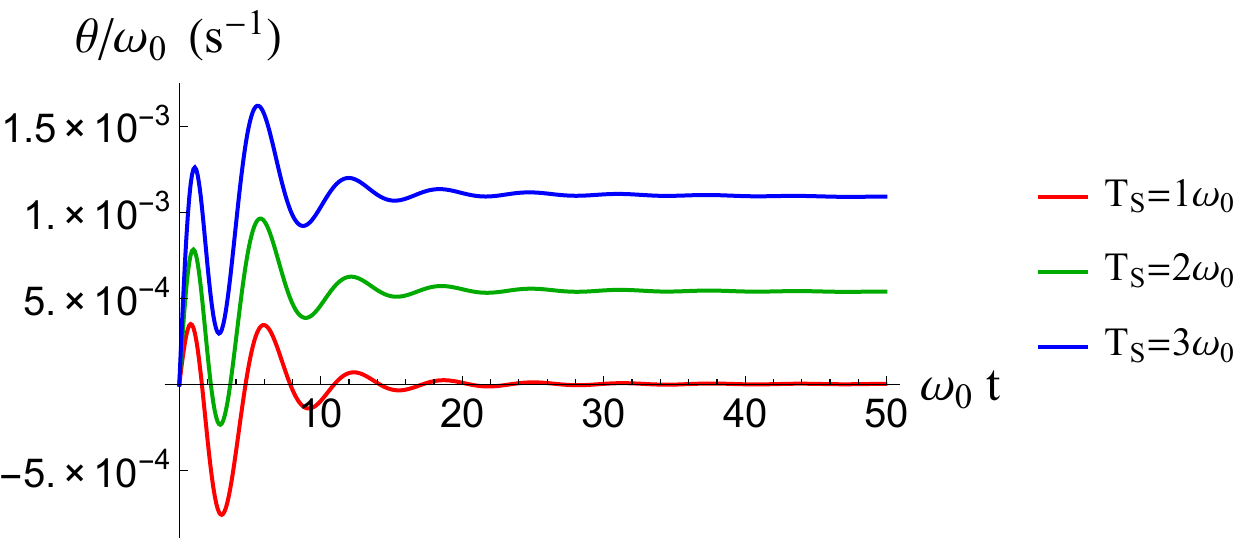}\\
{\bf (b)}\\
 \includegraphics[width=0.95\columnwidth]{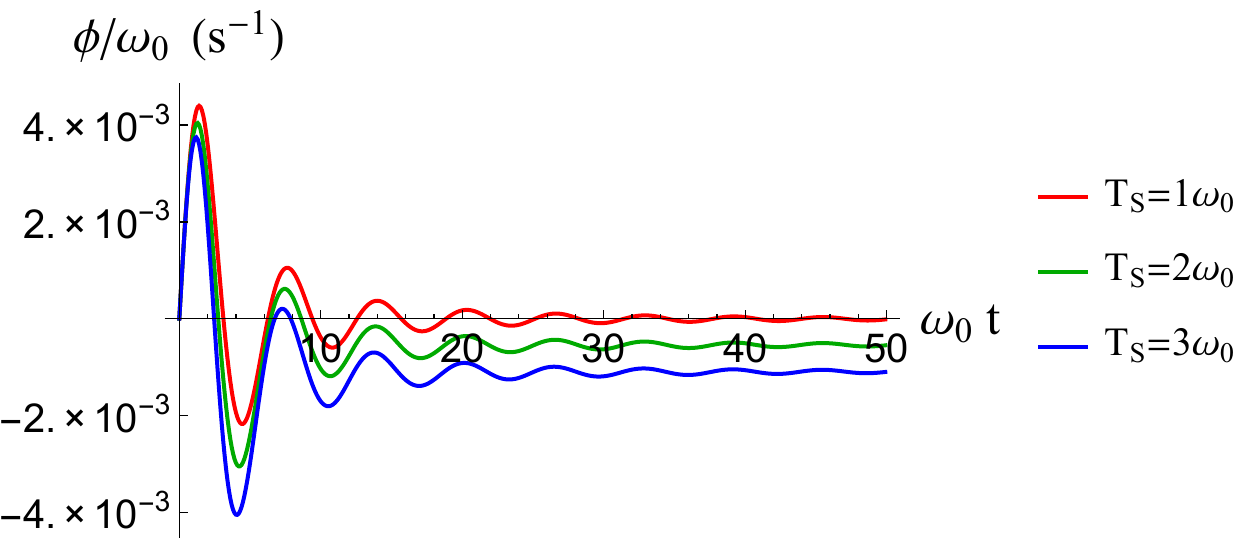}\\
\caption{(Color online) Time evolution of: (a) $\theta(t)$ and (b) $\phi(t)$, in units of $\omega_0$, for $\Omega=0.25\omega_0 $, $\lambda=0.01$ and $T_E=\omega_0$ and for different values of the initial system's temperature $T_S/\omega_0=1,\,2,\,3$.}
\label{fig1}
\end{center}
\end{figure}

In Fig. \ref{fig1}(a) we show the energy flow per unit of time $\theta(t)$ as given by Eq. \eqref{thetaFINAL}, in the weak coupling limit $\lambda=0.01$ and in units of $\omega_0$, for $\Omega=0.25\omega_0$, $T_E=\omega_0$ and for three different values of the effective system's temperature $T_S/\omega_0=1,\,2,\,3$.

An interesting feature of the energy flow is represented by the first positive peak of $\theta(t)$, which can be observed even when the initial temperatures of the reduced system and of the environment are equal to each other. Such peak, which was observed also in the case of a spin-boson model \cite{OurWork2015}, is a general feature due to the choice of dealing with an initial factorized state, which is essential in order to have a well-defined dynamical map \cite{Breuer2002}. In fact, even if system and environment are in Gibbs form relative to the same temperature $T$, i.e. $\rho_{SE}(0) = \frac{e^{-\Ham_S/T}}{Z_S} \otimes \frac{e^{-\Ham_E/T}}{Z_E}$,
with $Z_S$ and $Z_E$ being the partition functions of the reduced system and environment respectively, the state does not represent an equilibrium preparation  \cite{OurWork2015,AnkerholdPRB2014,SchmidtPRB2015}.
In particular, the contribution of the interaction Hamiltonian is absent before $t=0$, i.e. when the first measurement of the environmental energy in the two-time measurement protocol outlined above is performed.
\Brev{The switching on of the interaction term results in a net energy flow both into the environment and into the system that takes place at the early stage of the coupled evolution.} In fact, if we compute the change in the system's energy 
\begin{equation}\label{ensys}
\mean{\Delta E_S}_t \equiv \mathrm{Tr}_S\left[\Ham_S \left(\rho_S(t) - \rho_S(0)\right)\right] = \sigma(0,t) - \sigma(0,0),
\end{equation}
we still observe a first positive peak in its time-derivative $\phi(t) \equiv \frac{d}{dt}\mean{\Delta E_S}$, as shown in Fig. \ref{fig1}(b) \cite{SchmidtPRB2015}. The last equality in Eq. \eqref{ensys} has been obtained by noting that $\mean{\Ham_S}_t = \frac{1}{2}\mean{X^2+P^2}_t \equiv \sigma(0,t)$.

\Brev{It is also interesting to consider the time behavior of the change in the mean values of the energies of the environment $\mean{\Delta q}_t$ Eq. \eqref{enmeanchar} and of the system $\mean{\Delta E_S}_t$ Eq. \eqref{ensys}. While the latter is always a positive quantity, it turns out that the energy of the environment, for different values of cut-off frequency and initial temperatures, shows in the weak coupling regime a decrease over time with respect to its initial value, given $T_E=T_S$. 
This lower energy value persists in the long time limit. 
Being in the weak coupling regime, we can assume the final state of
the composite system to be effectively factorized, with an
environmental reduced density matrix that can therefore be cast into a
Gibbs form relatively to an inverse temperature which is lower than
the initial one. In this sense one could speak of a non-externally induced \textit{cooling effect}.}

\begin{figure}[htbp!]
\begin{center}
{\bf (a)}\\
\includegraphics[width=0.8\columnwidth]{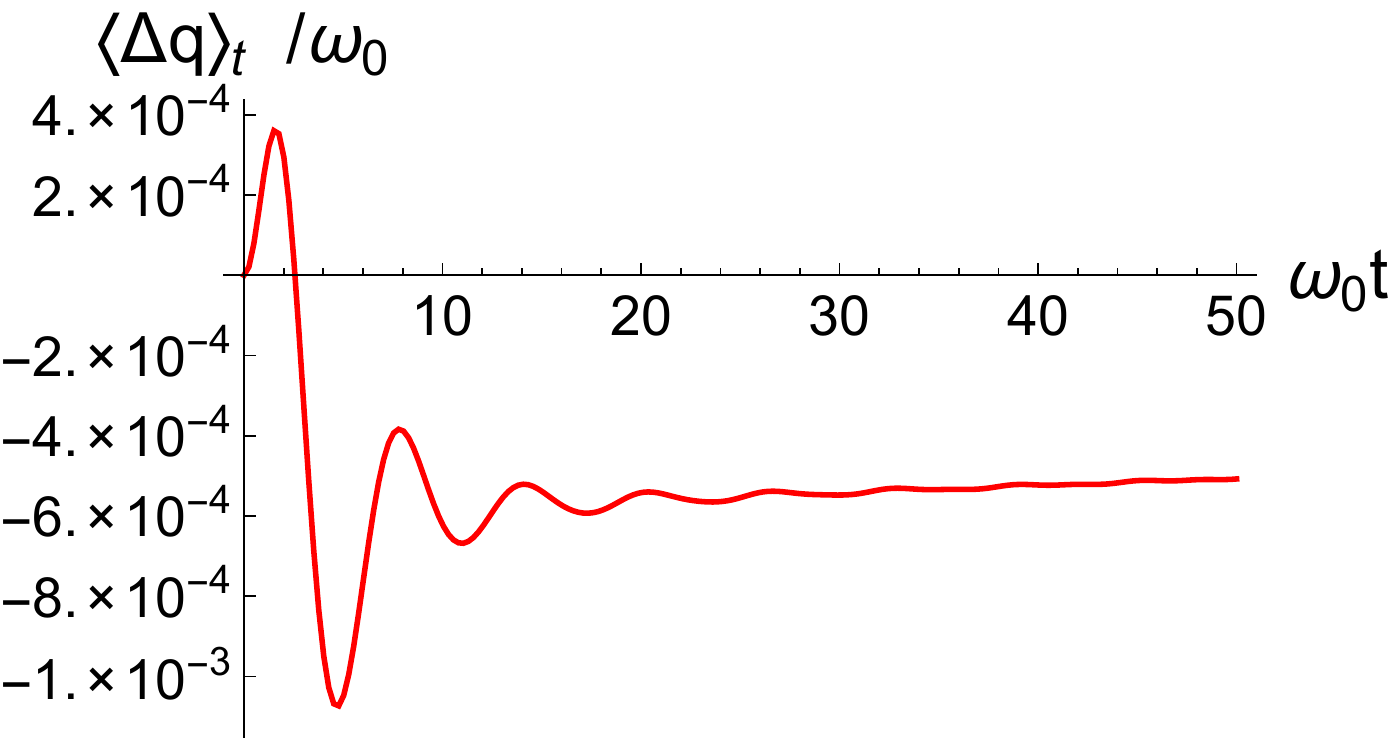}\\
{\bf (b)}\\
 \includegraphics[width=0.8\columnwidth]{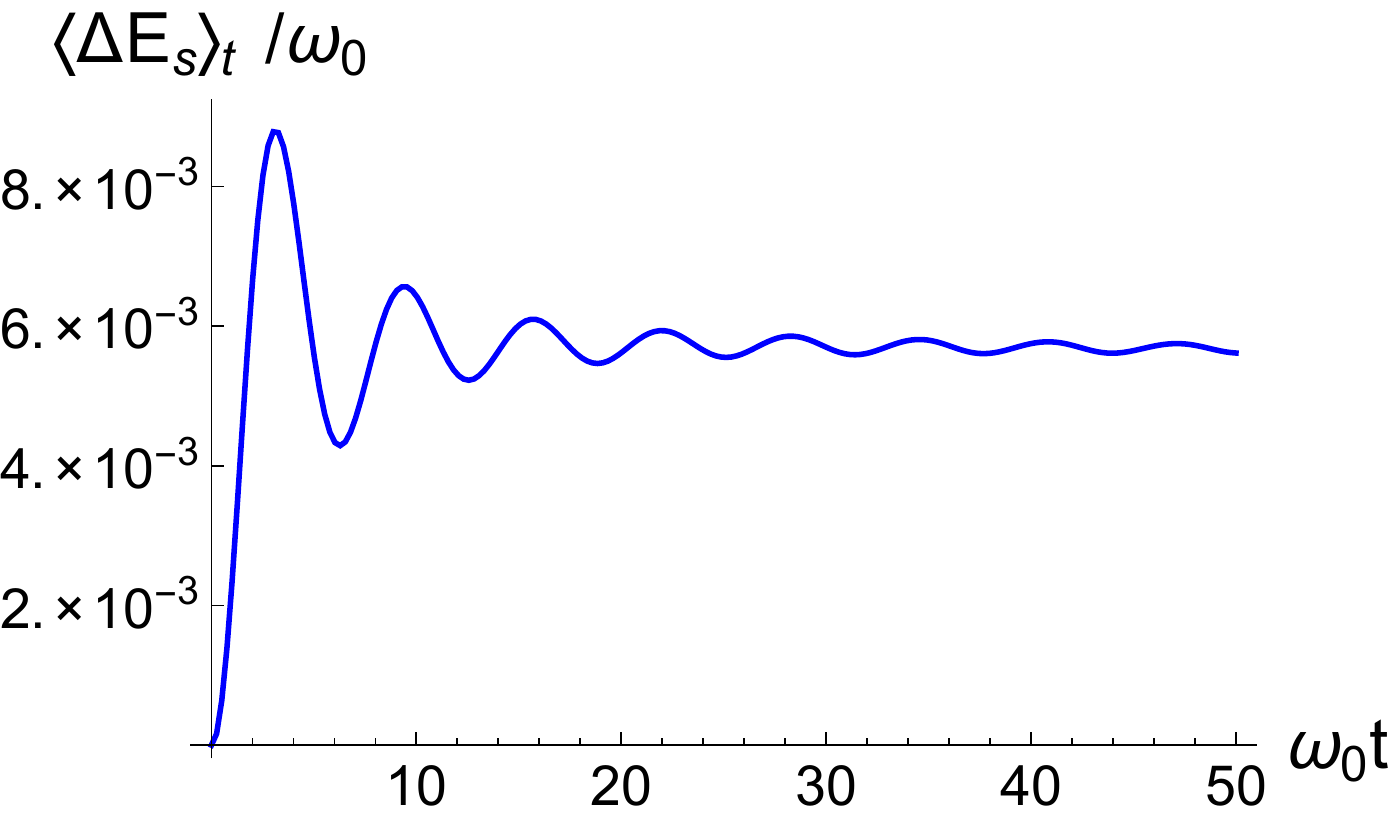}\\
\caption{(Color online) Time evolution of: (a) $\mean{\Delta q}_t$ and (b) $\mean{\Delta E_S}_t$, in units of $\omega_0$, for $\Omega=0.25\omega_0 $, $\lambda=0.01$ and $T_E=T_S=\omega_0$. Note that the final value of the internal energy of the environment is lower than its initial value, meaning that the environment has cooled down.}
\label{fig2}
\end{center}
\end{figure}

Finally, from the analysis of Fig. \ref{fig1}(a), it emerges how the energy backflow measure, i.e. the area of the negative region of $\theta(t)$, is maximized for $T_E = T_S$; strong numerical evidences suggest that this trend is maintained for all values of the relevant parameters $\lambda,\Omega,T_E$. This fact, in agreement also with what happens in the case of a spin-boson model \cite{OurWork2015}, can be understood considering that there is no initial temperature gradient when the two temperatures initially match, this favouring a more symmetric situation of energy exchange.
Exploiting this result, we can then evaluate the amount of energy backflow, as estimated by Eq. \eqref{enbackflow}. 

\begin{figure}[htbp!]
\begin{center}
{\bf (a)}\\
\includegraphics[width=0.95\columnwidth]{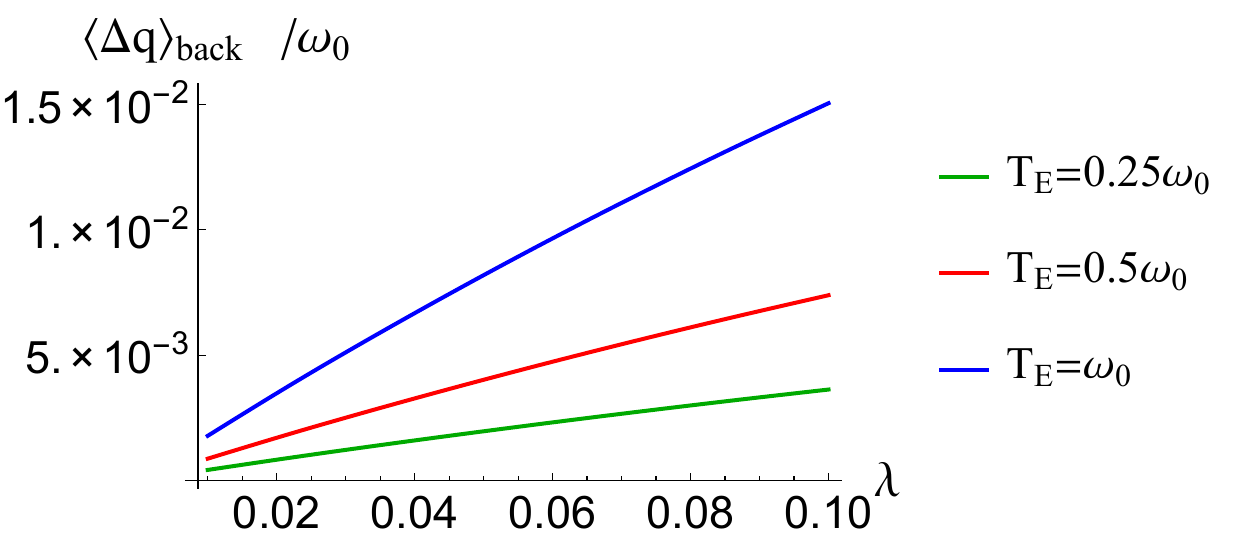}\\
{\bf (b)}\\
\includegraphics[width=0.95\columnwidth]{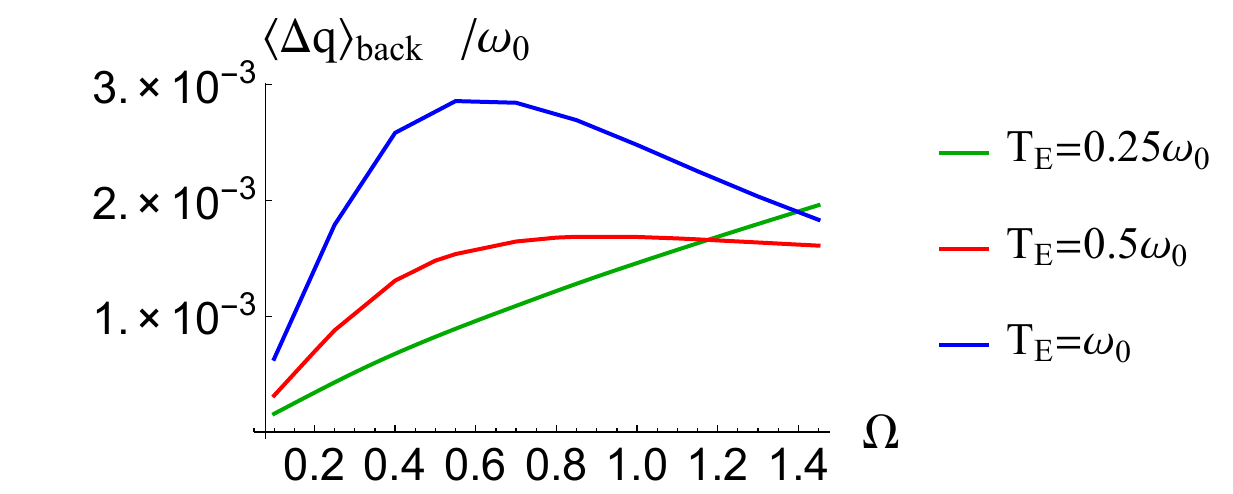}
\caption{(Color online) Energy backflow measure as a function of the coupling strength $\lambda$ for different values of the parameters $\Omega$ and $T_E$ which characterize the dynamical map.}
\label{fig3}
\end{center}
\end{figure}

In Fig. \ref{fig3} we show the behaviour of the energy backflow measure $\mean{\Delta q}_{back}$ with respect to its dependence on the various parameters $\lambda,\, \Omega,\, T_E(=T_S)$ in the range $\lambda\in\left[0.01,\,0.1\right]$. In such weak coupling regime the measure turns out to be monotonically increasing with the coupling strength and possesses a non trivial behavior with respect to the cut-off frequency $\Omega$: for intermediate values of the initial temperatures $T_E=T_S$, $\mean{\Delta q}_{back}$ decreases for large $\Omega$, while for very low temperature ($T_E=T_S=0.25\omega_0$) there seems to be an almost linear increment of the latter with $\Omega$.

\subsection{Energy backflow in the strong coupling regime}
\label{sec:Num}

\GGrev{In this Section we present a numerical approach to calculate
  the quantities $\mean{\Delta q}_{back}$ and $\theta(t)$,
  defined in Eqs.\eqref{enbackflow} and \eqref{thetachar}
  respectively, in the QBM without relying on the
  FCS. The results obtained this way will encompass both the dynamical
  regimes of weak coupling, where we will show the agreement with
  those obtained in the previous Sec. \ref{sec:QBMan}, and of strong coupling.}

The starting point of this method \cite{Vasile} is to consider the environment as composed by a large but \textit{finite} number $N$ of bosonic modes, so that the total Hamiltonian \eqref{Ham} of system+environment is now given in terms of a $ (N+1)\times (N+1)$ matrix of the form
\begin{equation}\label{HamNum}
\Ham = \frac{\mathbf{P}^T\vett{P}}{2} + \mathbf{X}^T \mathbf{M}\mathbf{X},
\end{equation} 
with $\mathbf{X} = \left( X_1, X_2,\ldots ,X_N, X_{N+1} \right)^T$ and $\mathbf{P} = \left( P_1, P_2,\ldots ,P_N, P_{N+1} \right)^T$ are the quadrature vectors and where the matrix $\mathbf{M}$ has elements $\mathbf{M}_{i,i} = \omega_i^2/2$ for $i = 1,\ldots , N$, $\mathbf{M}_{N+1,N+1} = \omega_0^2/2$ and $\mathbf{M}_{i,N+1} = \mathbf{M}_{N+1,i} = -g_i/2$, $g_i$ being the couplings between the system and the environmental mode $i$. 

The exact evolution for the position and momentum operators can be
formally written by exploiting the diagonalization of $\Ham$ (see Appendix \ref{app:B}).
The result reads 
\begin{align} \label{ExNumEqs}
& X_i (t) = \sum_{j=1}^{N+1} \left[  \mathbf{M}^{XX}_{ij}(t) X_j(0) + \mathbf{M}^{XP}_{ij}(t) P_j(0)   \right] \notag\\
& P_i (t) = \sum_{j=1}^{N+1} \left[  \mathbf{M}^{PX}_{ij}(t) X_j(0) + \mathbf{M}^{PP}_{ij}(t) P_j(0)   \right] \notag\\
\end{align}
where $ \mathbf{M}^{XX} (t) = \mathbf{M}^{PP} (t) \equiv \mathbf{O}\, \mathbf{Cos}\, \mathbf{O}^T $, $ \mathbf{M}^{XP} (t) \equiv \mathbf{O} \,\mathbf{Sin}\, \tilde{\mathbf{D}}^{-1}\, \mathbf{O}^T $ and finally $ \mathbf{M}^{PX} (t) \equiv \mathbf{O} \,\mathbf{Sin}\, \tilde{\mathbf{D}} \,\mathbf{O}^T $. 
In these expressions $\mathbf{O}$ denotes the orthogonal transformation which diagonalizes the total Hamiltonian $\Ham$, i.e. such that $\mathbf{M} = \mathbf{O}\mathbf{D}\mathbf{O}^T$ with $\mathbf{D} = \mathrm{diag}\left(\sqrt{2d_i}\right)$, while $\mathbf{Cos}, \mathbf{Sin}$ and $\tilde{\mathbf{D}}$ are diagonal matrices with elements $\mathbf{Cos}_{i,i} = \cos\left(d_i t\right)$, $\mathbf{Sin}_{i,i} = \sin\left(d_i t\right)$ and $\tilde{\mathbf{D}}_{i,i} = d_i $.

We stress that Eq. \eqref{ExNumEqs} requires no assumption but the finite number of harmonic oscillators. This gives rise to a different evolution at very long times (longer the higher is $N$), when the dynamics in the case of the finite environment leads to Poincar\'{e} revivals. 
However, since no weak coupling or secular approximations are involved in this exact numerical approach, it is possible to extend our study of energy backflow for this model also to the strong coupling regime $\lambda > 0.1 $, while confronting the numerical evidences with the analytical predictions in the weak coupling regime.

\begin{figure}[htbp!]
{\bf (a)}\\
\includegraphics[width=0.8\columnwidth]{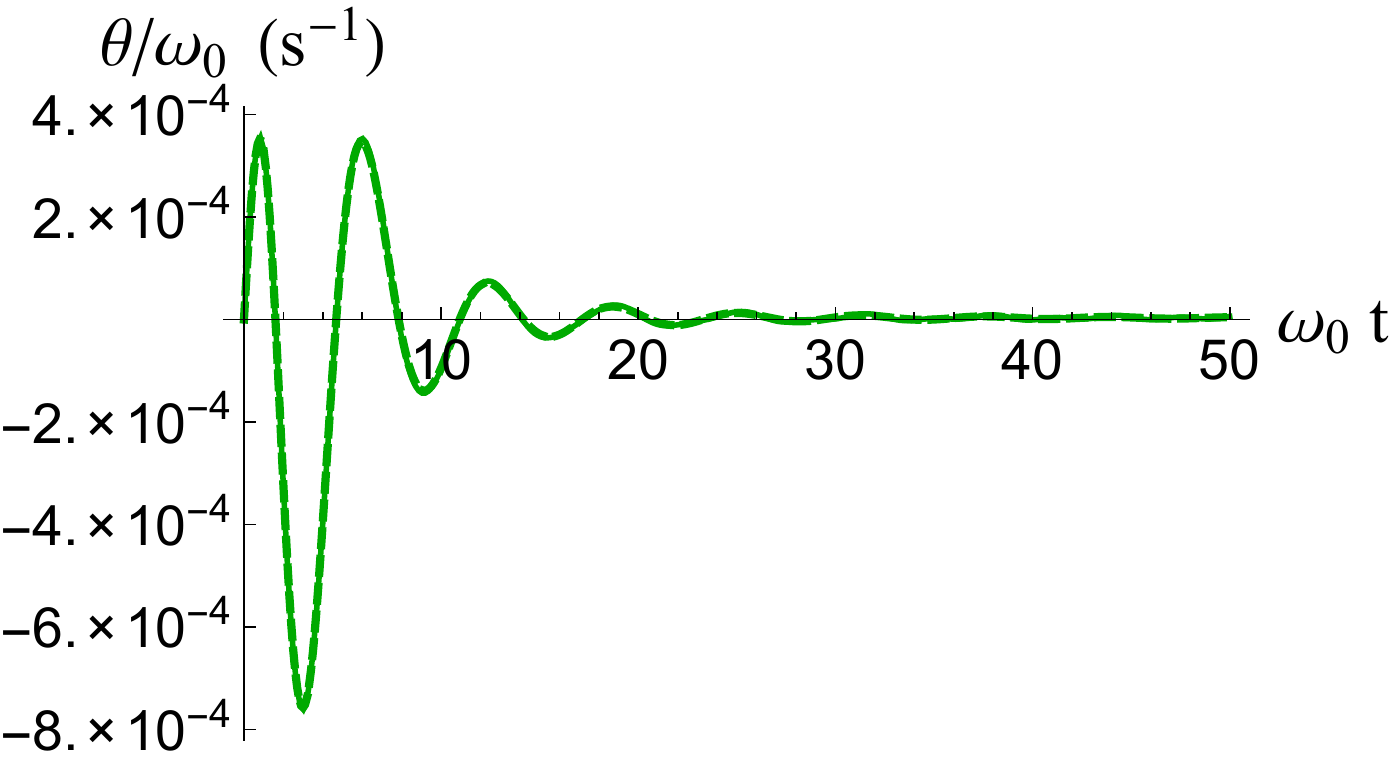}\\
{\bf (b)}\\
\includegraphics[width=0.8\columnwidth]{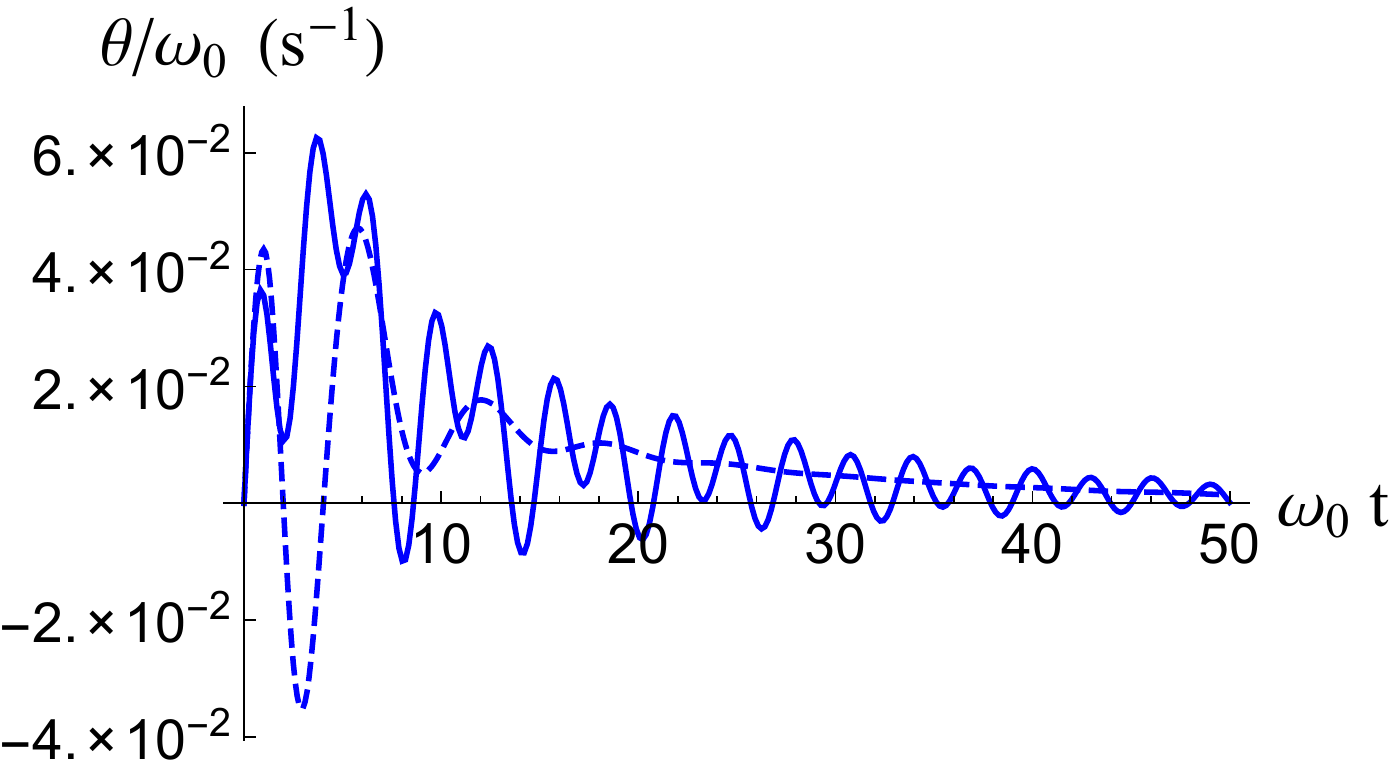}\\
{\bf (c)}\\
\includegraphics[width=0.8\columnwidth]{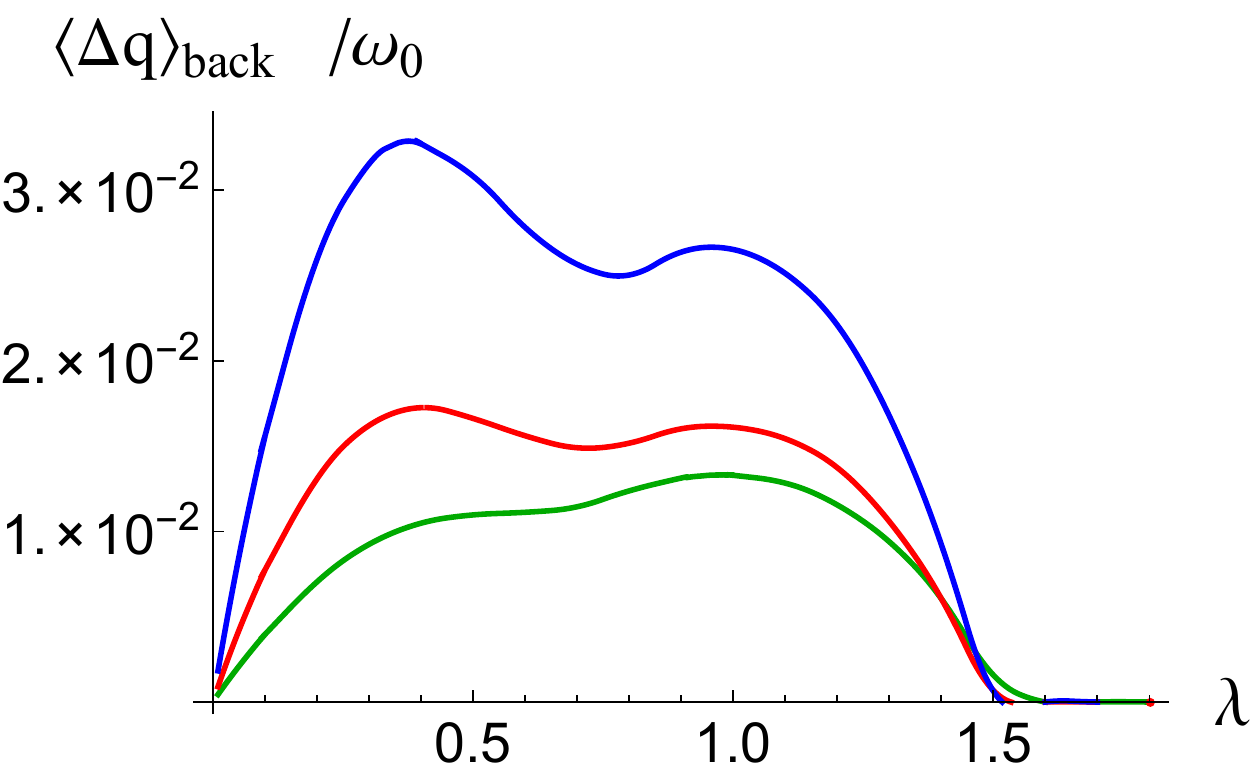}\\
{\bf (d)}\\
\includegraphics[width=0.8\columnwidth]{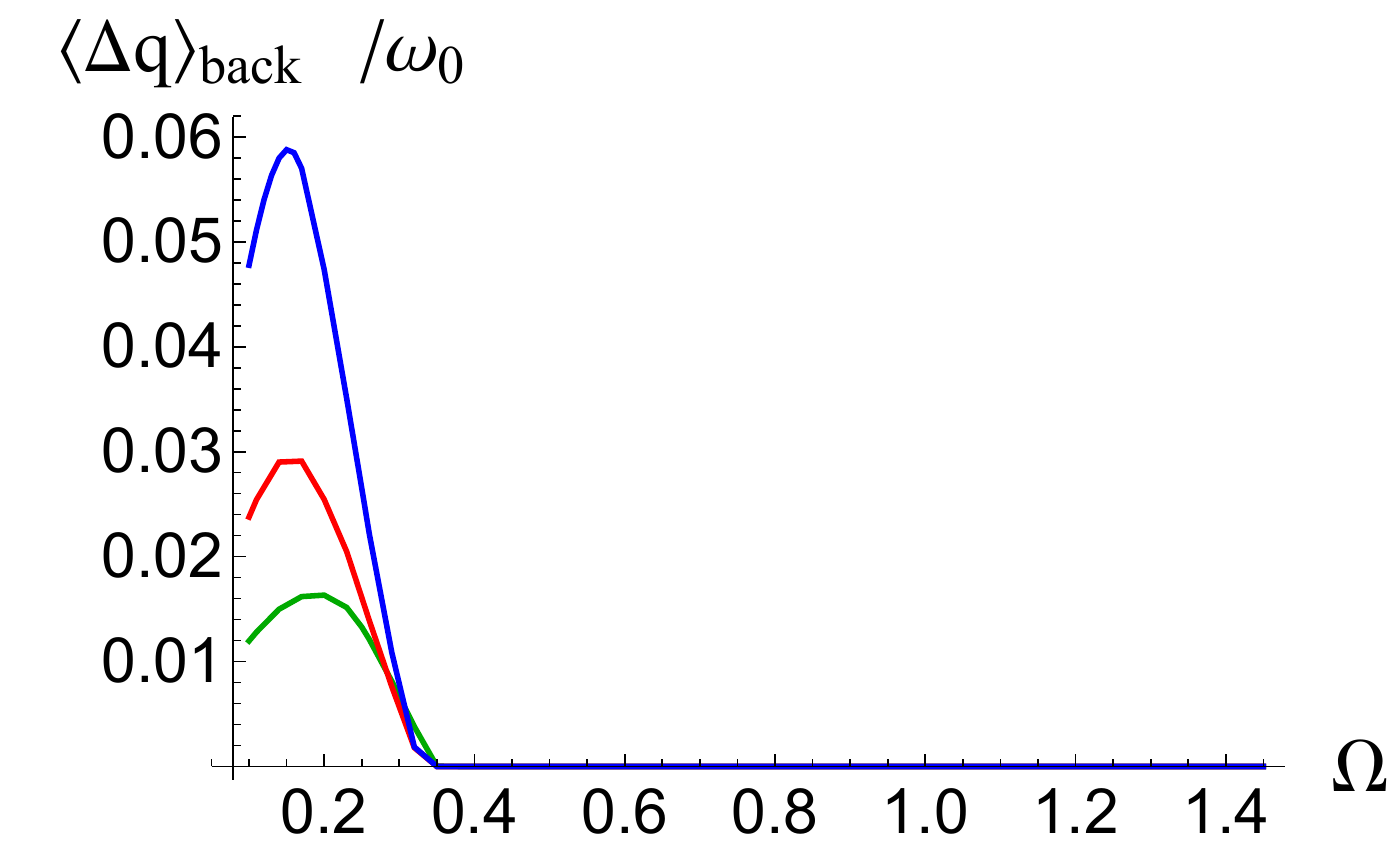}\\
\caption{(Color online) Time behavior of the energy flow per unit of time $\theta(t)$ in units of $\omega_0$ for $\Omega=0.25\omega_0$, $T_E=T_S=\omega_0$ and coupling strength $\lambda=0.01$ \textbf{(a)} and $\lambda=1$ \textbf{(b)}. The solid lines refer to the solution obtained with the numerical method, while the dashed lines are the curves predicted by the analytical approach relying on the FCS methods. Plot of the energy backflow measure in units of $\omega_0$ as a function of the coupling strength $\lambda$ for $\Omega=0.25\omega_0$ \textbf{(c)} and as a function of the coupling strength $\Omega$ for $\lambda = 1$ \textbf{(d)}, for three different values of the initial temperatures: $T_E = T_S = 0.25\omega_0$ (green line), $T_E = T_S = 0.5\omega_0$ (red line) and $T_E = T_S = \omega_0$ (blue line).
These curves were produced by means of the numerical simulation with $N=150$ environmental bosonic modes.}
\label{fig4}
\end{figure}

Building on Eq. \eqref{ExNumEqs}, we can straightforwardly obtain the energy flow per unit of time $\theta(t)$, which we plot in Fig. \ref{fig4}(a) having chosen $N=150$ modes in the environment for the simulation \cite{footnote2}. Fig. \ref{fig4}(a) clearly shows that in the weak coupling regime the numerical solution (solid line) retraces perfectly the predictions of the analytical approach based on the full-counting statistics (dashed line), while for strong coupling the difference between the two becomes marked, see Fig. \ref{fig4}(b).
Having $\theta(t)$ as a result of the numerical simulation and by means of Eq. \eqref{enbackflow}, it is then immediate to obtain the energy backflow measure, which we show as a function of $\lambda$ and of $\Omega$ in Figs. \ref{fig4}(c) and (d) respectively.
Note that the range of the coupling strength in Fig. \ref{fig4}(c), being $\lambda\in\left[0.01,\, 1.8\right]$, encompasses also the strong coupling regime; looking at the lower bottom-left corner, i.e. for $\lambda\in\left[0.01,\, 0.1\right]$, we can see that we recover the results obtained using the analytic approach shown in Fig. \ref{fig3}.

It is evident from Fig. \ref{fig4}(c) the existence of a threshold value of the coupling strength $\lambda^*\left(\Omega , T_E\right)$ above which the energy backflow measure vanishes. It can be shown that this behavior is maintained for any value of the cut-off frequency and temperature, proving therefore a general feature of the dynamics of this model.
In order to understand this result, we make use of Eq. \eqref{ExNumEqs} to calculate all the separate contributions to the total mean energy, i.e the time-evolution of the change in the mean values of the energy of the environment 
\[ 
\mean{\Delta q}_t = \frac{1}{2}\sum_{i=1}^N\left[\left(\mean{X_i^2}_t+\mean{P_i^2}_t\right)-\left(\mean{X_i^2}_0+\mean{P_i^2}_0\right)\right],
\]
of the system 
\begin{multline}
\mean{\Delta E_S}_t = \frac{1}{2}\left[\left(\mean{X_{N+1}^2}_t+\mean{P_{N+1}^2}_t\right)\right.\\
\left.-\left(\mean{X_{N+1}^2}_0+\mean{P_{N+1}^2}_0\right)\right],\notag
\end{multline}
and finally of the interaction Hamiltonian $\mean{\Delta \Ham_I}_t $.

\begin{figure*}[htbp!]
{\bf (a)}\hskip0.31\textwidth {\bf (b)} \hskip0.31\textwidth {\bf (c)}
\hspace*{-0.5cm}\includegraphics[width=0.31\textwidth]{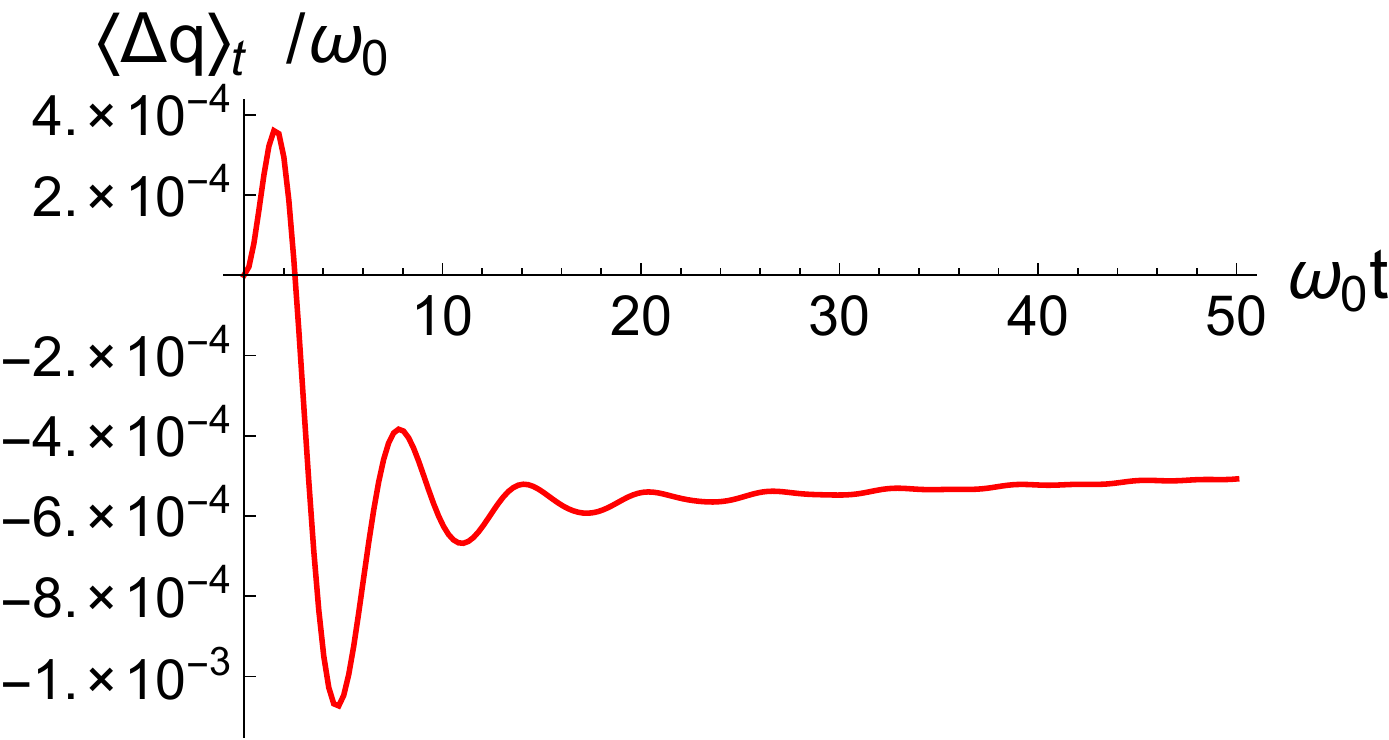}~\hspace*{0.5cm} \includegraphics[width=0.31\textwidth]{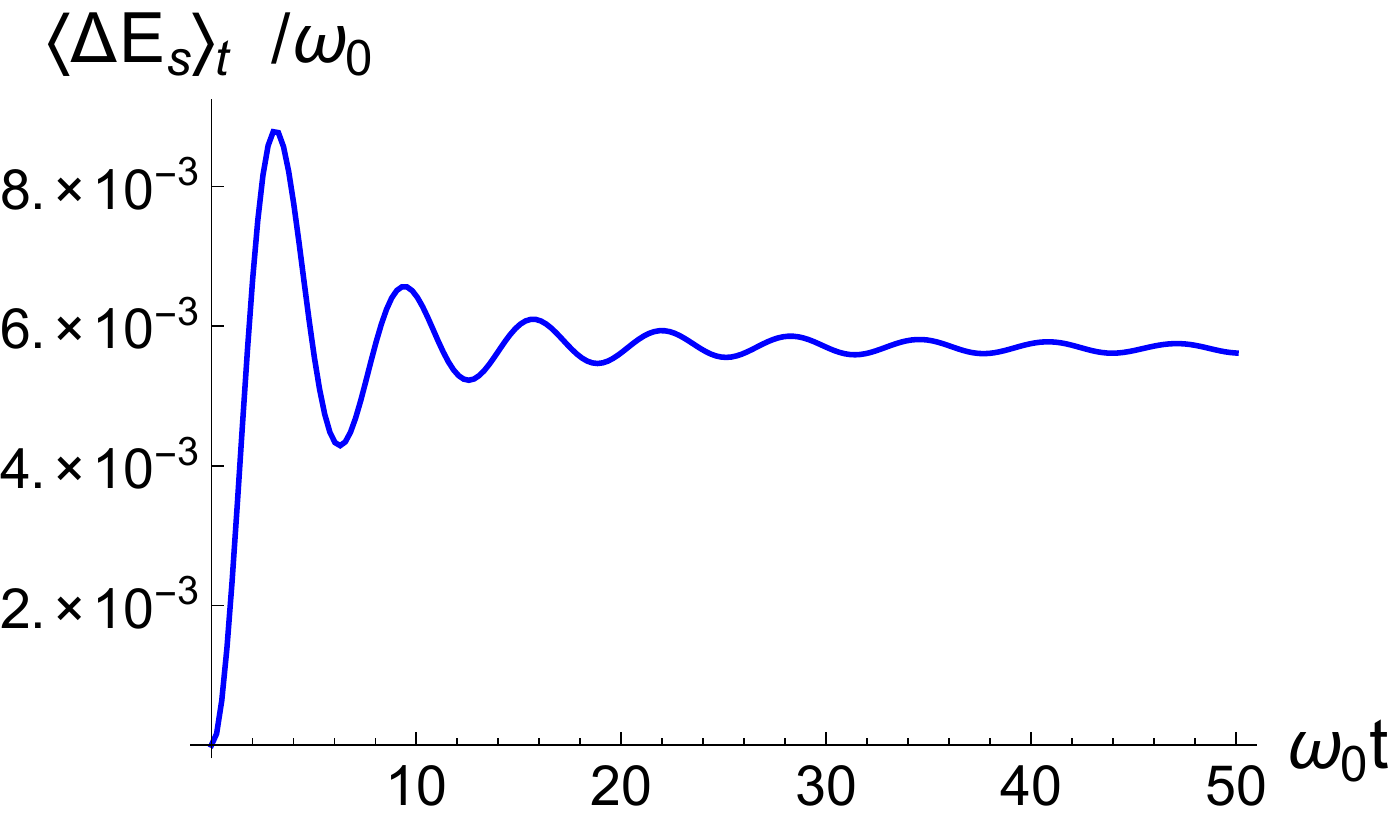}~
\includegraphics[width=0.31\textwidth]{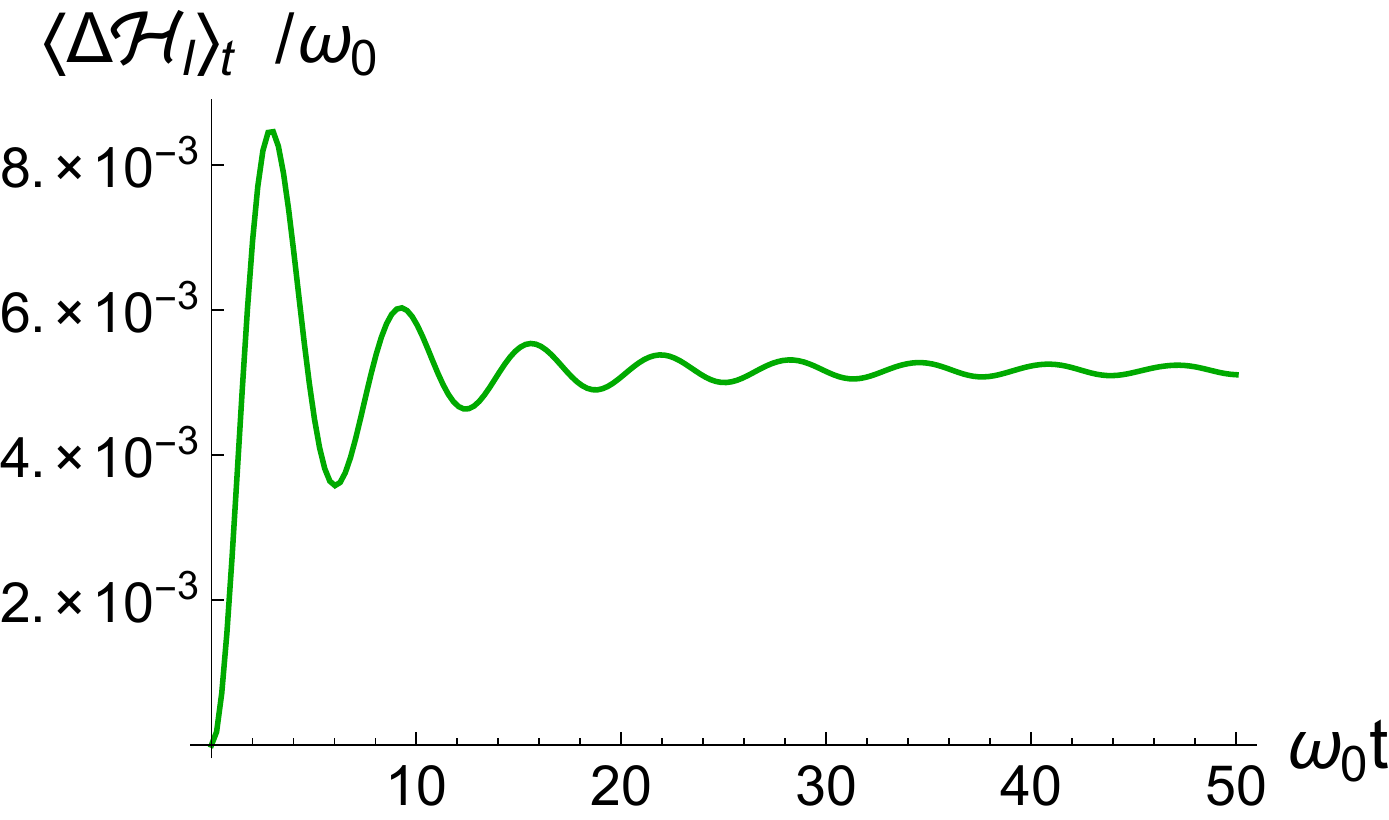}\\
{\bf (d)}\hskip0.31\textwidth {\bf (e)} \hskip0.31\textwidth {\bf (f)}
\hspace*{-0.2cm}\includegraphics[width=0.31\textwidth]{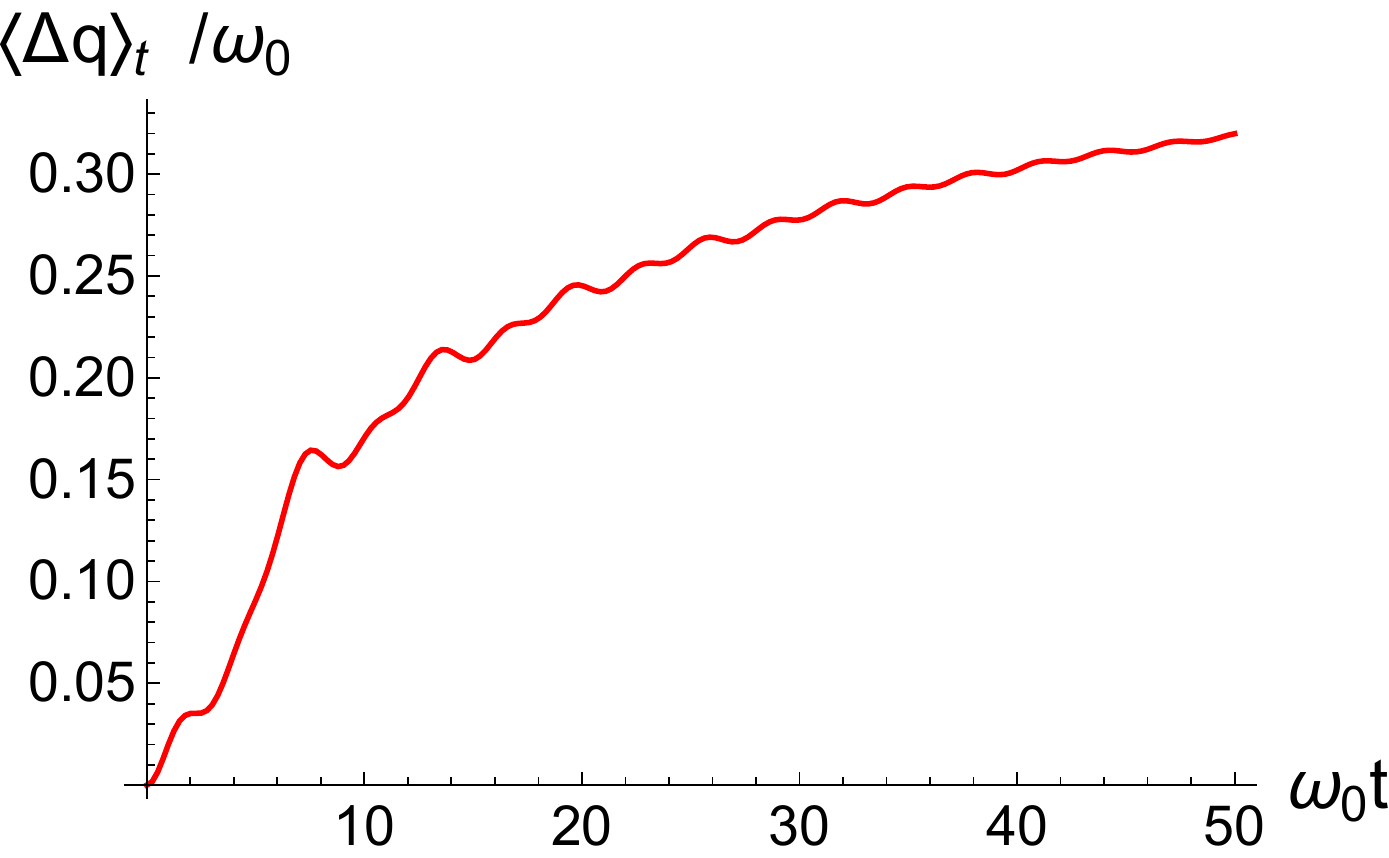}~\hspace*{0.1cm} \includegraphics[width=0.31\textwidth]{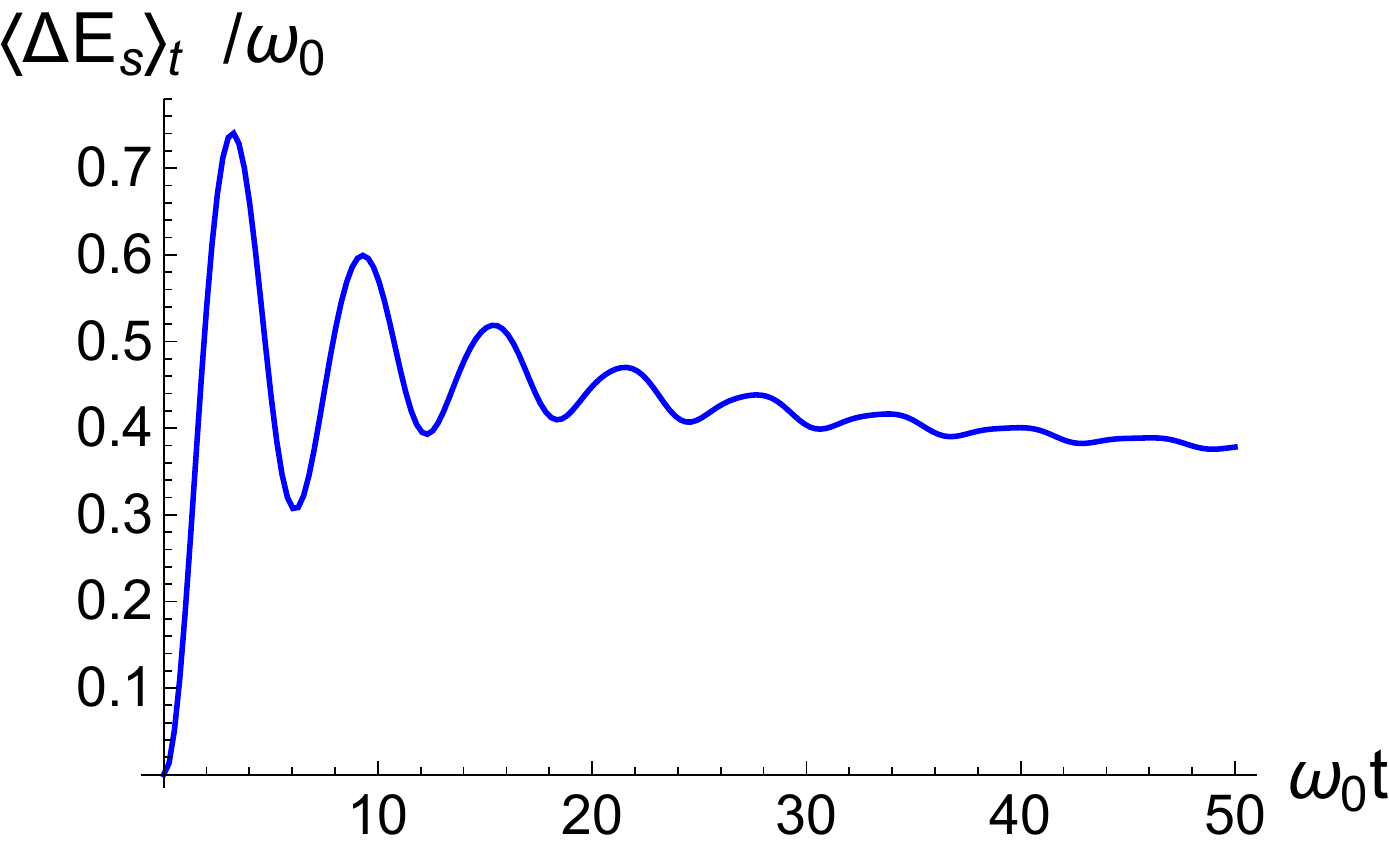}~\hspace*{0.3cm}
\includegraphics[width=0.31\textwidth]{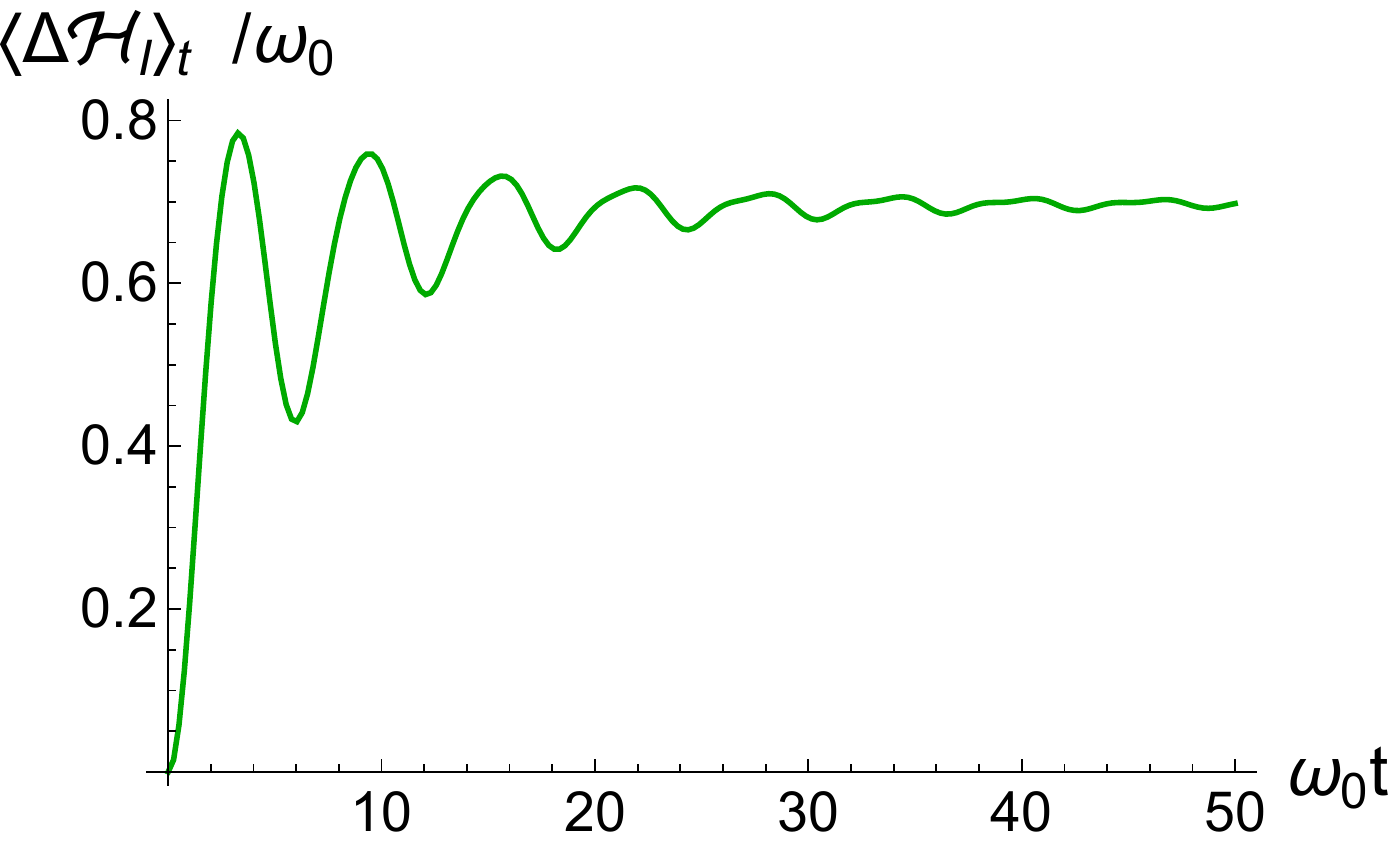}\\
{\bf (g)}\hskip0.31\textwidth {\bf (h)} \hskip0.31\textwidth {\bf (i)}
\includegraphics[width=0.31\textwidth]{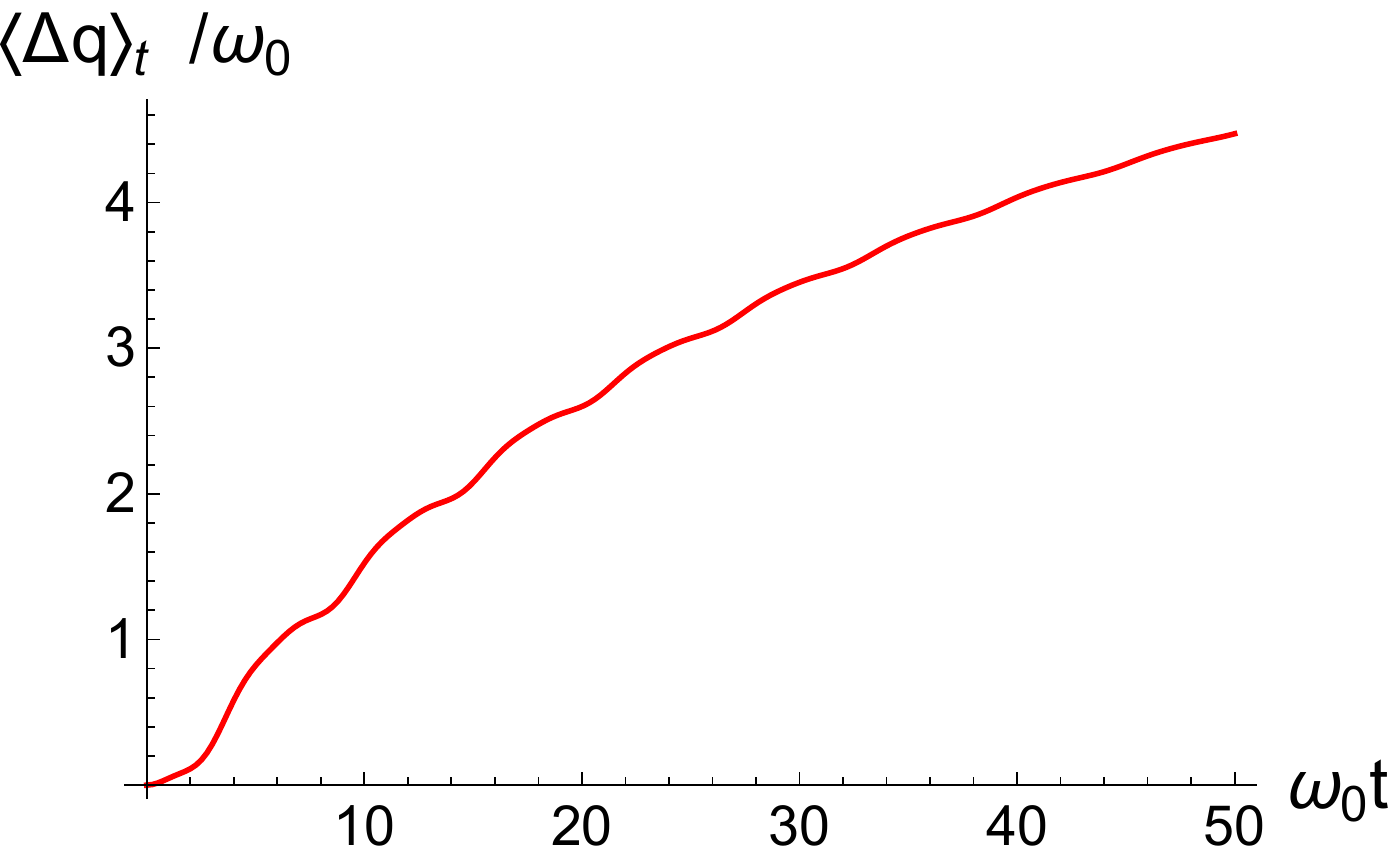}~ \includegraphics[width=0.31\textwidth]{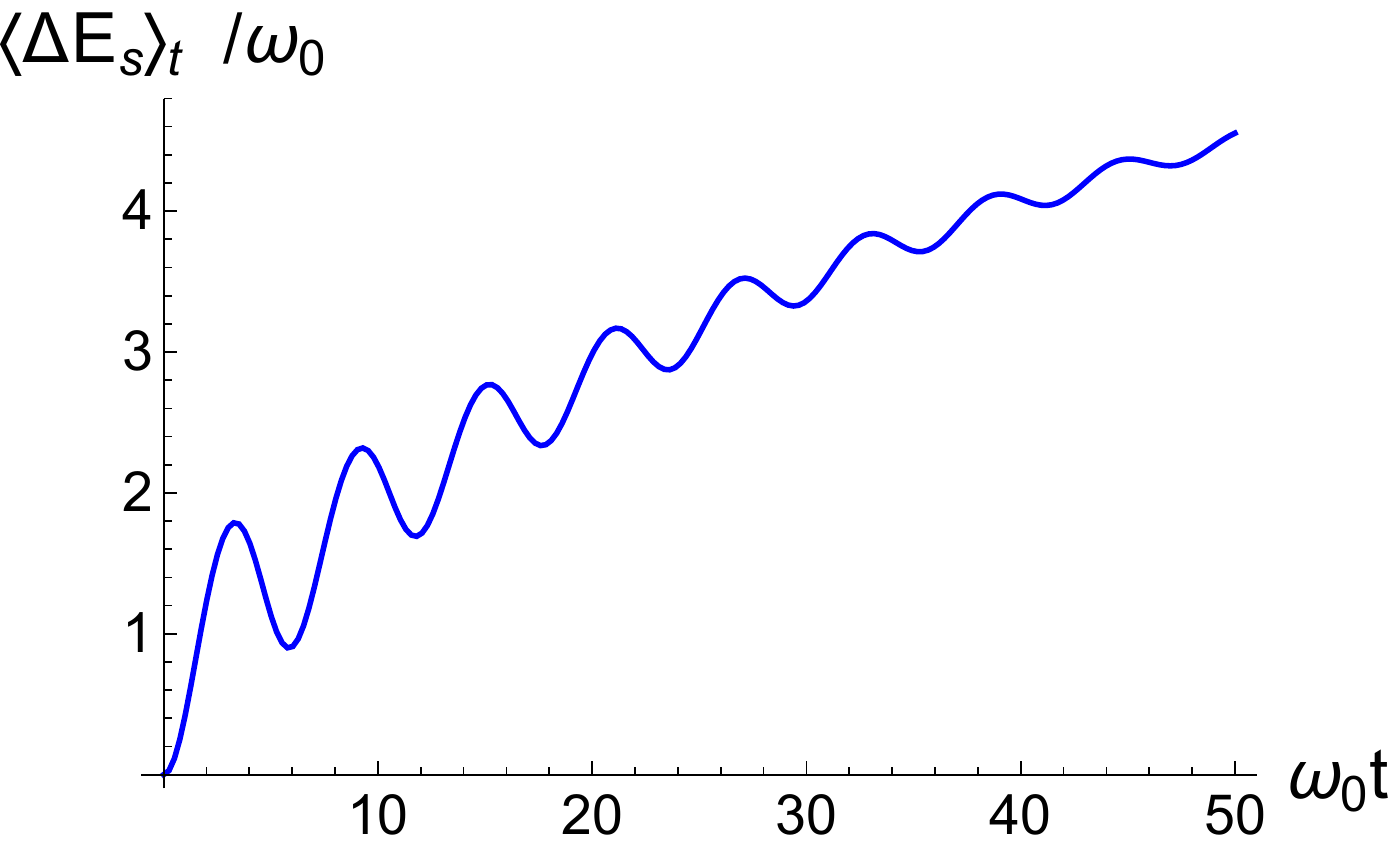}~
\includegraphics[width=0.31\textwidth]{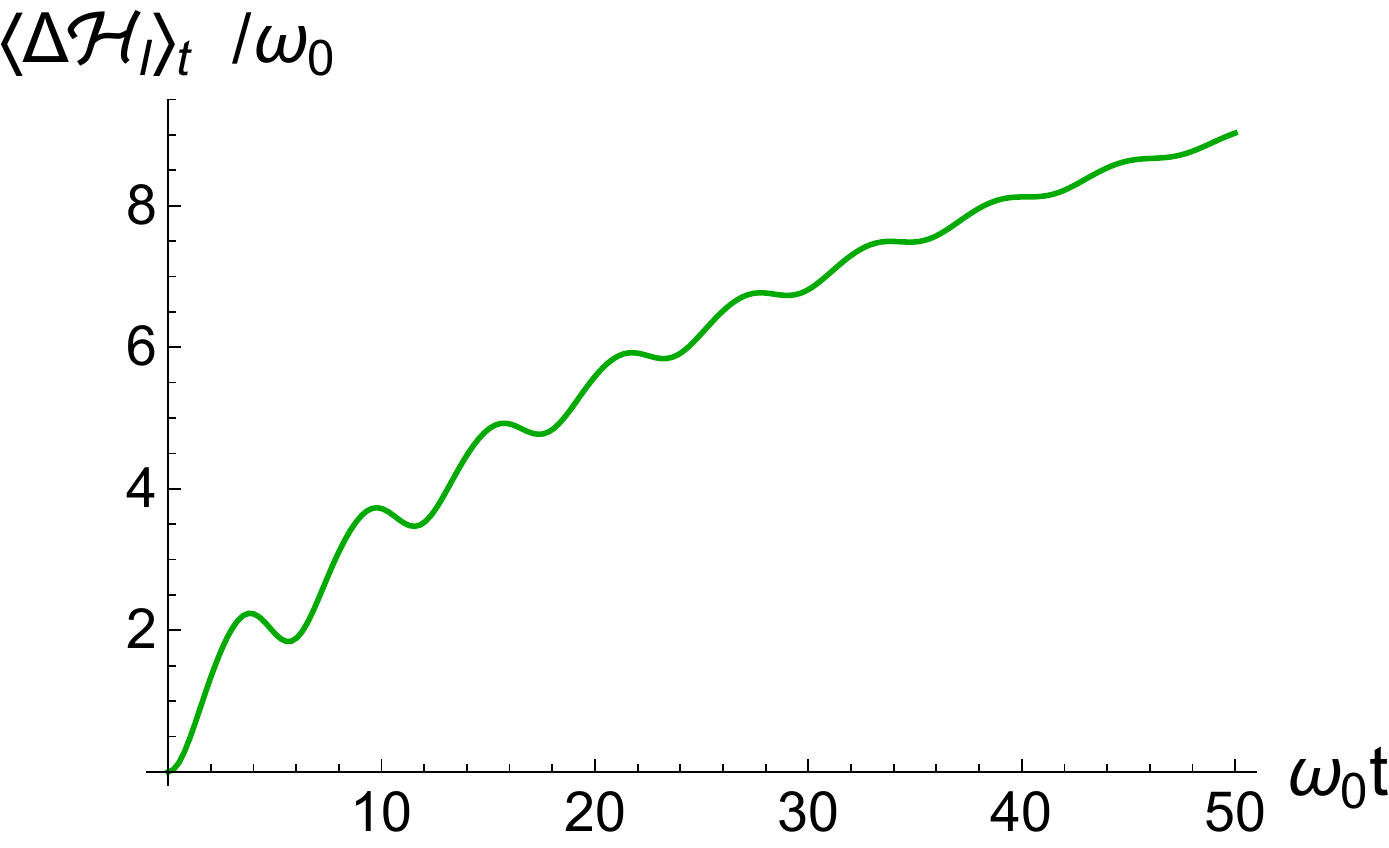}~
\caption{Separate contributions to the time behavior of the mean total energy in units of $\omega_0$, for $ \Omega=0.25\omega_0,\, T_E=T_S=\omega_0$. The top three panels (a-c) refer to the weak coupling case $\lambda=0.01$, the middle three (d-f) to $\lambda=0.8$ and finally the bottom three (g-i) to strong coupling $\lambda=1.8 > \lambda^*$. (a), (d) and (g) Mean values of the change in the environmental Hamiltonian Eq. \eqref{enmeanchar}; (b), (e) and (h) Mean values of the change in the system Hamiltonian Eq. \eqref{ensys}; (c), (f) and (i) Mean values of the change in the interaction Hamiltonian.}
\label{fig5}
\end{figure*}

Fig. \ref{fig5} shows these three different contributions for $\Omega=0.25\omega_0$ and $T_E=T_S=\omega_0$ in the cases of weak coupling $\lambda=0.01$ (top three plots) and of strong coupling $\lambda=0.8$ (middle three plots) and $\lambda=1.8$ (bottom three plots), the latter corresponding to a situation for which the energy backflow measure vanishes.
The energy backflow contributions correspond to the time regions where the mean internal energy of the environment [red curves in panels (a), (d) and (g)] temporarily decreases. The measure introduced in Eq. \eqref{enbackflow} is just the sum of all these contributions.

Fig. \ref{fig5} shows the different time behavior of the average energy of the environment [panels (a), (d) and (g)]. In particular, in the weak coupling regime the latter decreases, see Fig. \ref{fig5}(a), this leading to the cooling effect previously put into evidence using FCS methods.
An opposite behavior is observed in the strong coupling, where the change in the average energy of the environment increases with time, see Fig. \ref{fig5}(d) and (g). For strong coupling the three contributions become of the same order of magnitude, at variance with what happens in the weak coupling case, where the change in the system's internal energy and in the mean value of the interaction Hamiltonian were roughly an order of magnitude bigger than the change in the environmental energy.
An analysis of these two cases shows then that in the weak coupling the time-variation of $\mean{\Delta E_S}_t$, which is always positive in our setup, is due both to 
the switching on of the interaction Hamiltonian at $t=0^+$ (after the energy measurement  on the environment in the two-time measurement protocol) but also to the backflow of energy from the environment, which, despite at the same initial temperature, loses to it a part of its energy.
In the strong coupling regime this no longer happens, and the increment in the mean system's energy is only due to $\mean{\Delta\Ham_I}_t$, which becomes dominant and ceases energy also to the environment, thus opposing the occurrence of energy backflow which is in fact very much reduced and eventually, when the threshold coupling strength $\lambda^*(\Omega, T_E)$ is reached and overcome (bottom three panels), stops.

\section{Relationship with the non-Markovianity of the reduced dynamics}
\label{sec:Markov}

We conclude the present work by studying the parameter dependence of the non-Markovianity in this QBM setting and comparing it with the behaviour of the energy backflow. To this purpose, we calculate a recently introduced measure of non-Markovianity \cite{Adesso1}, based on the time behaviour of the Gaussian Interferometric Power (GIP). Employing the quantum Fisher information, the GIP measures the ability to estimate, according to black-box interferometry, a local phase shift in a worst case scenario with a two-mode Gaussian probe \cite{Adesso2, Bera} characterizing the state of the reduced system plus an ancilla. The GIP is a measure of discord-type correlations between system $S$ and ancilla $A$ and can be calculated from the symplectic invariants of the joint covariance matrix $\bgreek{\sigma}_{SA}$. It is monotonically non-increasing under local completely - positive and trace preserving maps acting on the reduced system. In the same spirit as  many other non-Markovianity measures \cite{Breuer2009PRL,Rivas2010PRL,Lu2010PRA,
LuoPRA2012,Lorenzo2013PRA,Bylicka2014SciRep,Chruscinski2014PRL}, non-Markovian dynamics are defined as those which lead to a non-monotonic behavior of the
GIP, i.e., such that there exist time intervals where
\begin{equation}
\mathscr{D}(t) \equiv \frac{d}{dt}\mathcal{Q}_G \left(\bgreek{\sigma}_{SA}\right) > 0.
\end{equation}
While the non-Markovianity measure $\mathscr{N}_{\mathcal{Q}}(\Lambda)$ \cite{Adesso1} includes a maximization over all possible initial two-mode Gaussian states:
\begin{eqnarray}
\mathscr{N}_{\mathcal{Q}}(\Lambda) &=& \max_{\bgreek{\sigma}_{SA}} \mathscr{N}^{\bgreek{\sigma}}_{\mathcal{Q}}(\Lambda)\nonumber\\
\mathscr{N}^{\bgreek{\sigma}}_{\mathcal{Q}}(\Lambda)&\equiv& \frac{1}{2}\int_0^{+\infty}\, dt\, \left(|\mathscr{D}(t)|+\mathscr{D}(t)\right),
\end{eqnarray}
$\mathscr{N}^{\bgreek{\sigma}}_{\mathcal{Q}}(\Lambda)$ represents a (more easily computable) lower bound for the latter. Analytic expressions for $\mathscr{N}^{\bgreek{\sigma}}_{\mathcal{Q}}(\Lambda)$ for the QBM in the weak coupling and secular approximation are given in \cite{Adesso2} for two important classes of initial two-mode Gaussian states: the mixed thermal states
(MTS) and the squeezed thermal states (STS), respectively characterized by covariance matrices of the form
\begin{equation}
\bgreek{\sigma}_{SA}^{MTS} = k e^{2r_1} \begin{pmatrix}
\mathbf{x}_1 & \mathbf{y}_1 \\
\mathbf{y}_1 & \mathbf{x}_1 \\
\end{pmatrix},\quad 
\bgreek{\sigma}_{SA}^{STS} = k \begin{pmatrix}
\mathbf{x}_2 & \mathbf{y}_2 \\
\mathbf{y}_2 & \mathbf{x}_2 \\
\end{pmatrix},
\end{equation}
where $\mathbf{x}_{1,2} = \mathrm{diag}(x_{1,2},x_{1,2})$ with $x_{1,2}=\cosh\left(2r_{1,2}\right)$ and where $\mathbf{y}_1 = \mathrm{diag}(y_1,y_1)$, $\mathbf{y}_2 = \mathrm{diag}(y_2,-y_2)$ with $y_{1,2}=\sinh\left(2r_{1,2}\right)$. In these expressions $k=\nu+1/2$, with $r_1$ being the strength of the Gaussian operations, $r_2$ the squeezing parameter and $\nu$ the average number of thermal photons.

\begin{figure}[htbp!]
\begin{center}
{\bf (a)} \hskip0.51\columnwidth {\bf (b)} 
\hspace*{-0.5cm}\includegraphics[width=0.45\columnwidth]{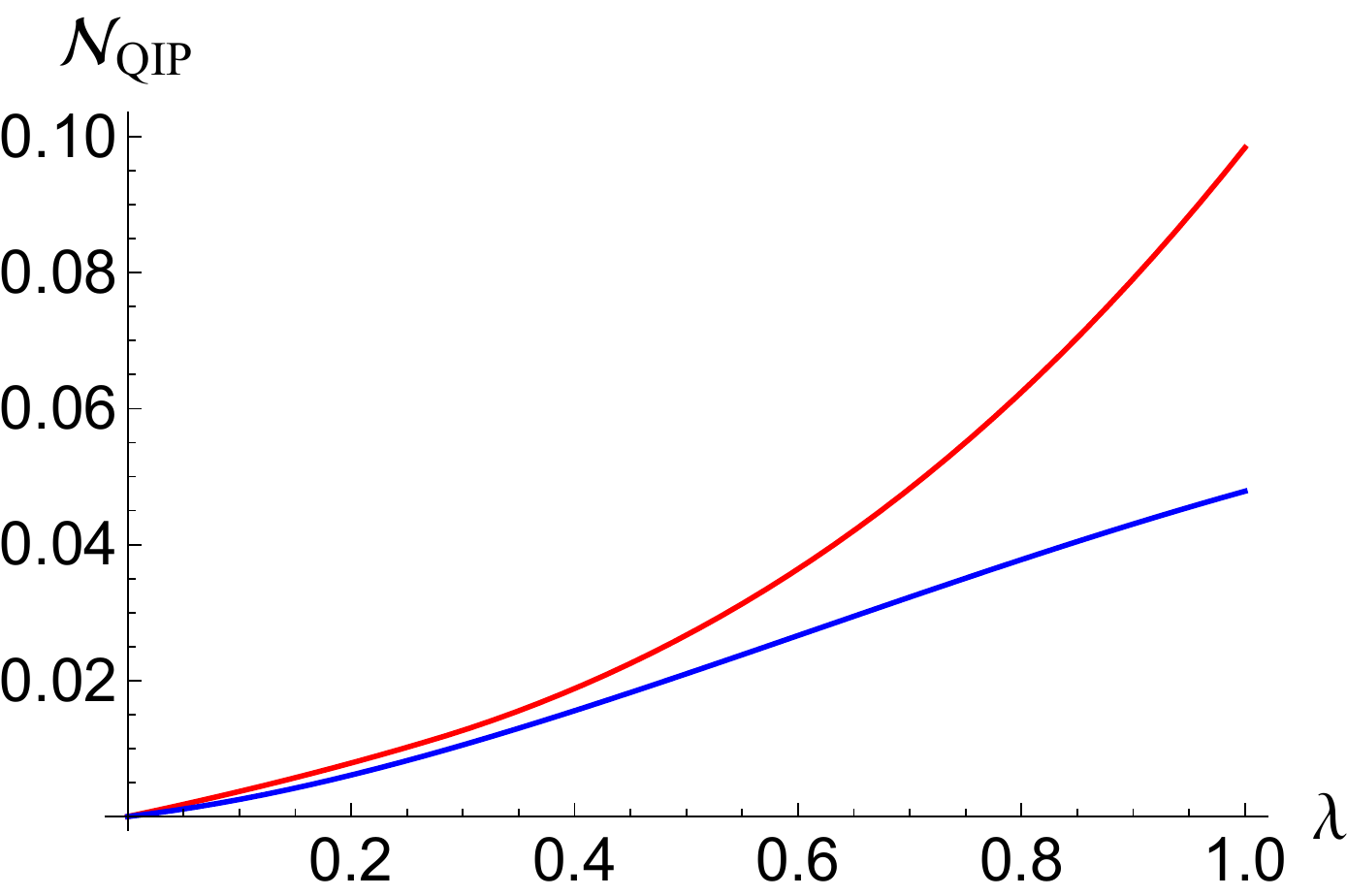}~
\includegraphics[width=0.45\columnwidth]{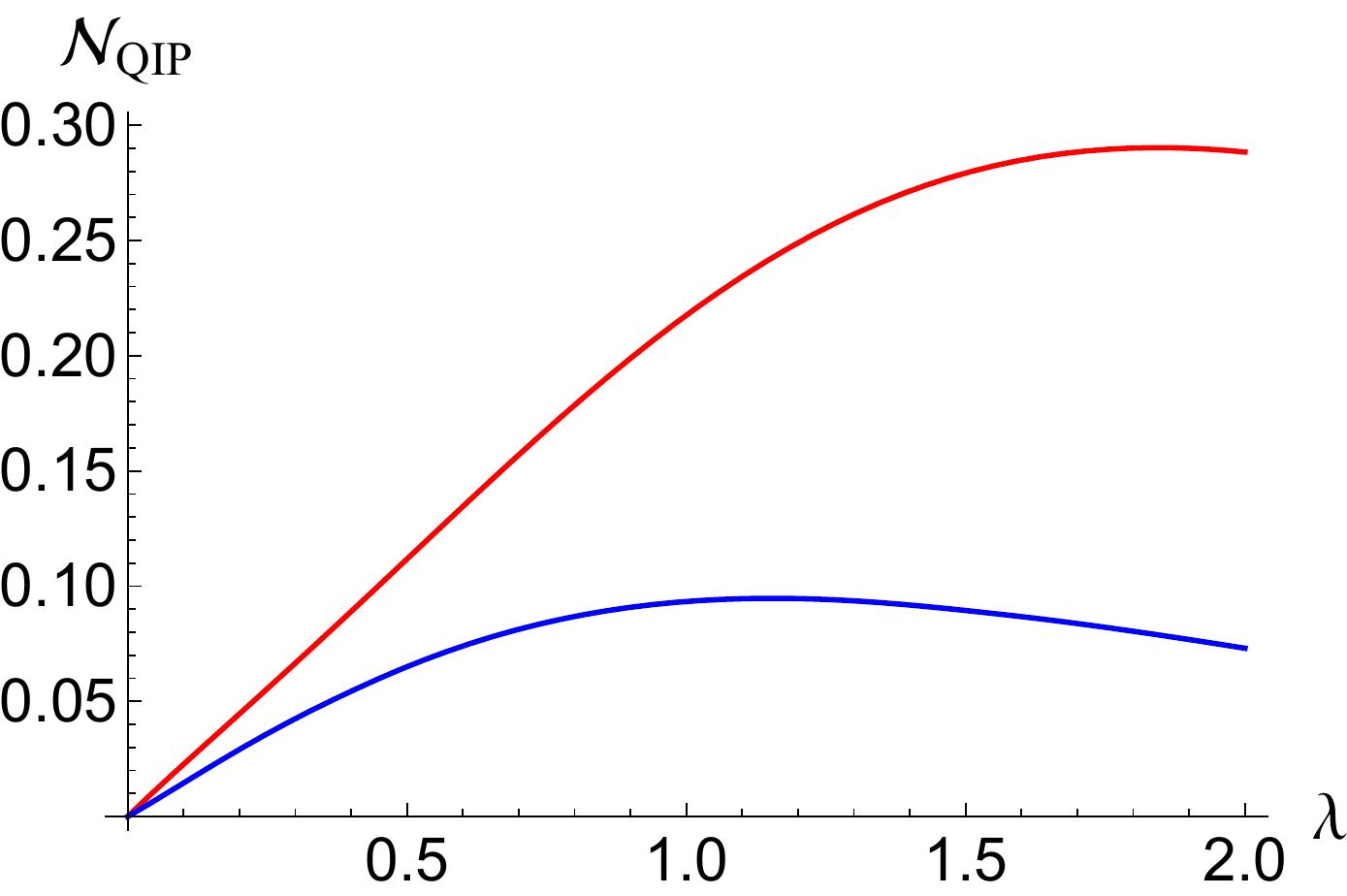}\\
{\bf (c)}\hskip0.51\columnwidth {\bf (d)}
\hspace*{-0.2cm}\includegraphics[width=0.45\columnwidth]{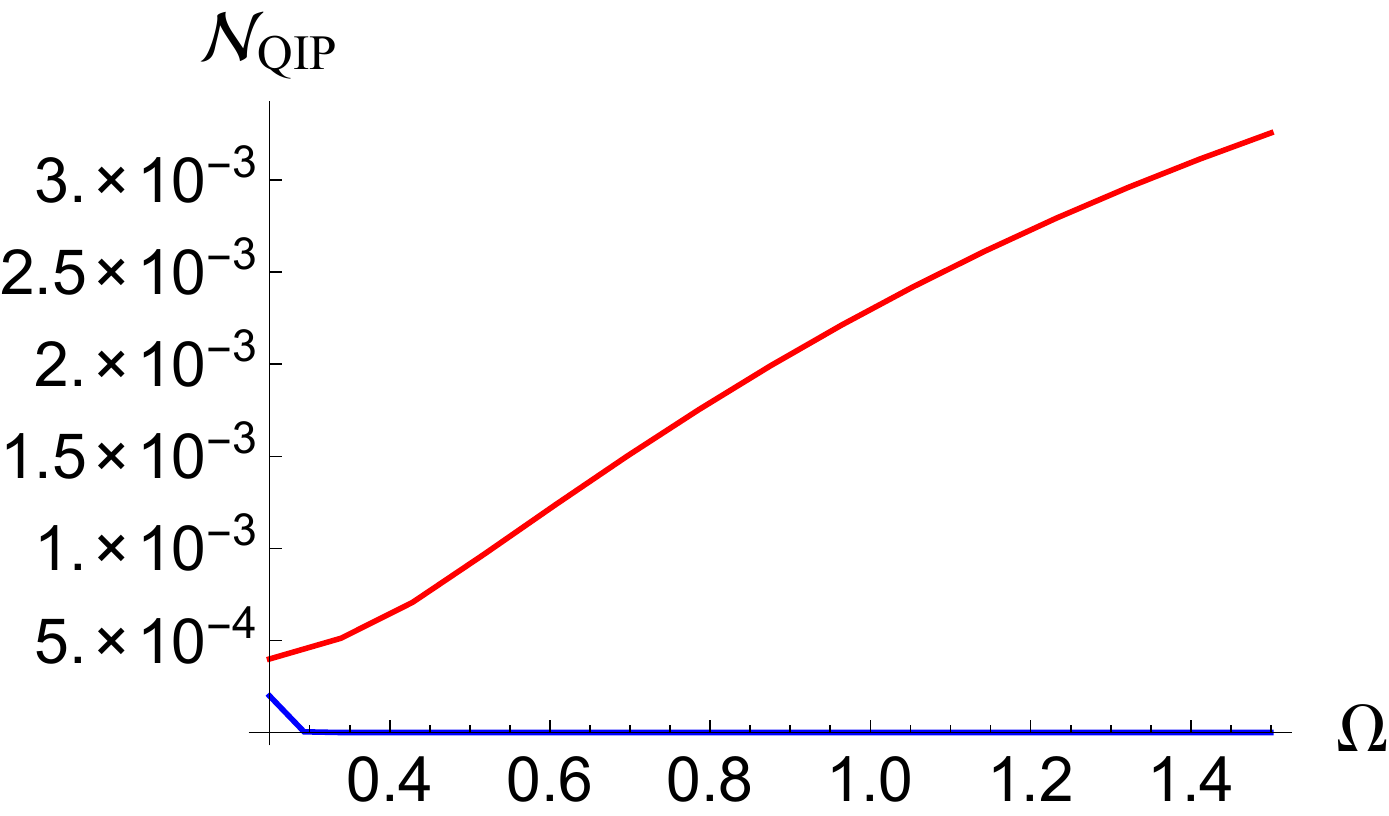}~\hspace*{0.1cm} \includegraphics[width=0.45\columnwidth]{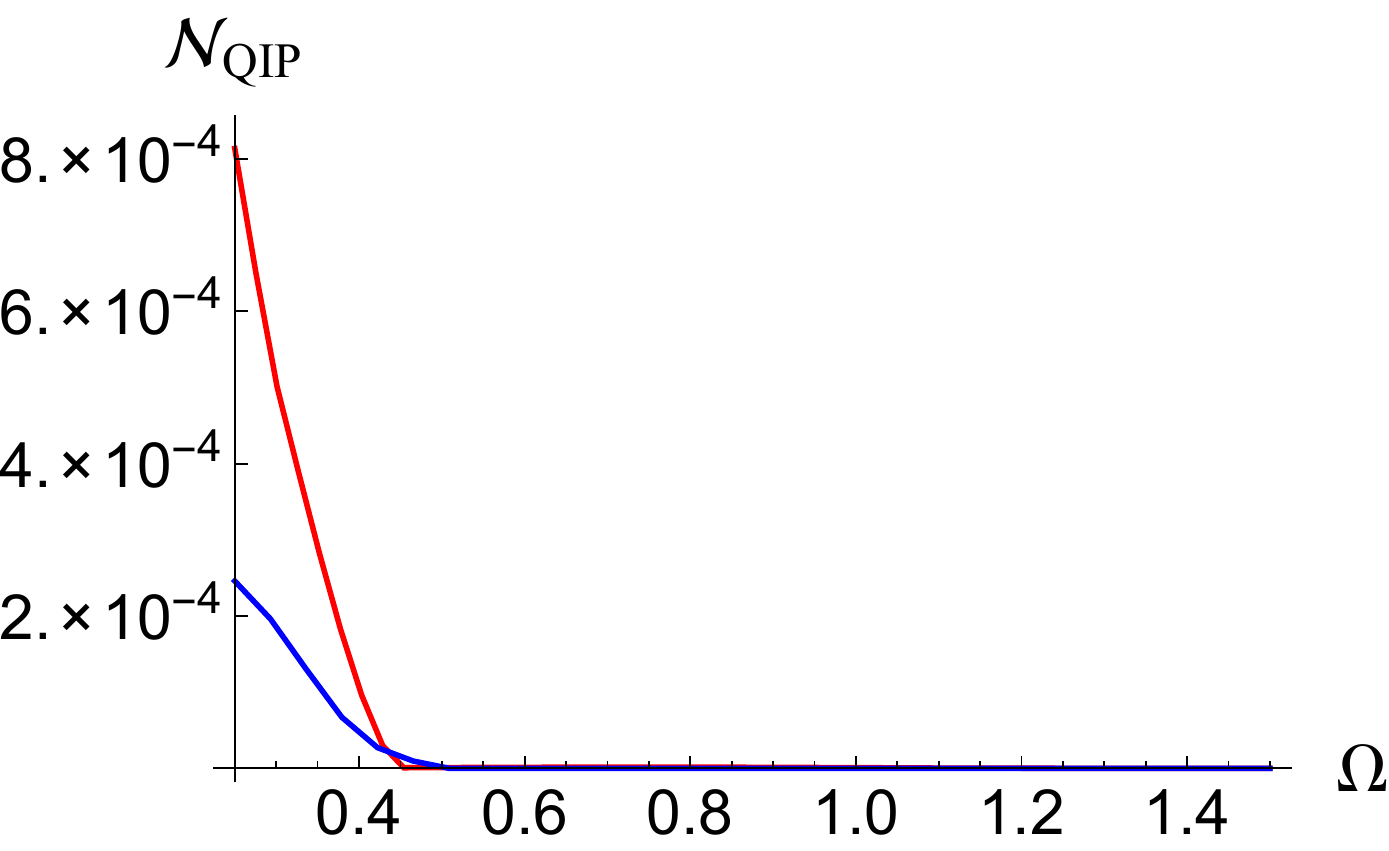}\\
\caption{(Color online) Plots of the non-Markovianity measure $\mathscr{N}^{\bgreek{\sigma}}_{\mathcal{Q}}$ for the class of STS (blue lines) and of MTS (red lines), with $k_{1,2} = 1$, $r_1 = r_2 = 0.658$ (otherwise stated), as function of $\lambda$ [panels (a) and (b)] and of the cut-off frequency [panels (c) and (d)] for fixes values of the remaining parameters. In particular: (a) $\Omega = 0.25\omega_0$ and $T_E=0.25\omega_0$; (b) $\Omega = 0.25\omega_0$ and $T_E=\omega_0$; (c) $\lambda=0.01$ and $T_E=0.25\omega_0$ and $r_1=10^{-2}$; (d) $\lambda=0.01$, $T_E=\omega_0$ and $r_1 = r_2 = 0.22$.}
\label{fig6}
\end{center}
\end{figure}

Fig. \ref{fig6} shows $\mathscr{N}^{\bgreek{\sigma}}_{\mathcal{Q}}$ as a function of the coupling strength $\lambda$ and as a function of the cut-off frequency $\Omega$ for fixed values of the remaining parameters $ T_E ,k$ and $r_{1,2}$. 
It turns out that the non-Markovianity measure is an increasing function of the coupling strength both for low and high temperatures, see Fig. \ref{fig6}(a) and (b), at variance with the energy backflow measure which shows a non-monotonic behavior on $\lambda$ and vanishes above a certain threshold $\lambda^*(\Omega,T_E)$.
The dependence on the cut-off frequency of the non-Markovianity measure and of the energy backflow measure is instead more similar: both these quantifier present in fact a monotonic increase with $\Omega$ in the low-temperature $T_E=0.25\omega_0$ and weak-coupling $\lambda=0.01$ regime, see Figs. \ref{fig6}(c) and green curve of Fig. \ref{fig3}(b), while for increasing values of the environmental temperature, both vanish above a certain value of the cut-off [see Figs. \ref{fig6}(d) and \ref{fig5}(d)].

\section{Conclusions}
\label{sec:Conclusions}

We have investigated the dynamics of the energy exchange in the QBM and characterized, through FCS formalism, the amount of energy which flows back from the environment to the reduced system in a generally non-Markovian setup.
To this purpose, we have pursued both an analytic approach in the weak coupling regime and a numerical method, valid for a large but finite number of bath modes, which has allowed to access also the strong coupling regime.
We also quantified the non-Markovianity of the reduced dynamics using a recently introduced witness based on the non-monotonicity of the Gaussian interferometric power.

In the weak coupling regime, the energy added by switching on the interaction, is fully transferred to the reduced system accompanied by a significantly smaller contribution from the environment. The latter results in a small reduction of the energy of the environment. For increasing values of the coupling strength, both reduced system and environment get their share of the interaction energy. Above a certain threshold value of coupling strength, which exhibits only a very weak dependence on the bath temperature, the energy backflow from the environment vanishes.

In perspective, it would be interesting to further deepen the investigation of the link between the occurrence of energy backflow and of memory effects in the reduced dynamics by relying on other suitable witnesses of non-Markovianity such as \cite{ParisManiscalco,Lorenzo2013PRA} and more structured thermal environments such as those described by sub- or super-Ohmic spectral densities \cite{Haikka2013PRA} or such as complex oscillator networks \cite{Johannes}.   

\acknowledgments
 
We gratefully acknowledge financial support by European Union (EU) through the Collaborative Projects QuProCS  (Grant Agreement 641277), by UniMI through the H2020 Transition Grant No. 14-6-3008000-623, the Academy of Finland through project no. 287750, the Centres of Excellence Programme (2015-2017) under project no. 284621, the Center of Quantum Engineering at Aalto University School of Science, the Jenny and Antti Wihuri foundation and the Magnus Ehrnrooth Foundation.

\appendix

\section{Calculation of the energy flow per unit of time $\theta (t)$ using FCS methods}

\label{app:A}

In this Appendix we explicitly show the calculations needed to derive result \eqref{thetaFINAL} using FCS methods.
As stated in Sec. \ref{sec:Formalism}, in order to calculate the energy flow per unit of time $\theta(t)$ and the consequent measure of energy backflow $\mean{\Delta q}_{back}$, we need to move to the phase-space representation and consider the characteristic function $\chi^{(\eta)}(\lambda,\lambda^*,t)$ defined in Eq. \eqref{charfunc}. The latter is obtained from the solution of the Fokker-Planck differential equation \cite{Gardiner} associated with the master equation \eqref{GenME1} for $\rho_S(\eta,t)$
\begin{widetext}
\begin{multline}\label{FokPl}
\frac{d}{dt}\chi^{(\eta)}(q,p,t) = \left\{\omega_0\left(q\partial_p - p\partial_q\right) - V_1(\eta,t) \left(\partial^2_{qq}+\partial^2_{pp}\right) - \left(2\Delta(t) + V_1(\eta,t)\right)\frac{q^2+p^2}{4}\right.\\
\left.+ \left(V_2(\eta,t) -\gamma(t) \right)\left(q\partial_q+p\partial_p\right) + V_2(\eta,t)\right\} \chi^{(\eta)}(q,p,t),
\end{multline}
\end{widetext}
where we have introduced the independent real variables $q = 2^{-1/2}(\lambda+\lambda^*)$, $p=i 2^{-1/2}(\lambda^*-\lambda)$ and the quantities $ V_{1,2}(\eta,t) = \frac{1}{2} \left(g_-(\eta,t) \pm g_+(\eta,t)\right) $.

Due to the quadratic nature of the Hamiltonian \eqref{Ham}, the Gaussian shape of the characteristic function is granted \cite{Carmichael, Olivares, Puri, Paternostro1} and thus an educated ansatz is
\begin{multline}\label{eq:ansatz}
\chi^{(\eta)}(q,p,t) = \Psi(\eta, t) \exp\left[i \,(q,p)^T \begin{pmatrix} X_m(\eta, t)\\P_m(\eta, t)\end{pmatrix} \right.\\
\left. - \frac{1}{2} \, (q,p)^T \begin{pmatrix}\sigma_{XX}(\eta, t) & \sigma_{XP}(\eta, t)\\\sigma_{PX}(\eta, t) & \sigma_{PP}(\eta, t)\end{pmatrix}\begin{pmatrix} q\\p\end{pmatrix}\right],
\end{multline}
where $ (X_m(\eta, t),P_m(\eta, t))^T \equiv (\mathrm{Tr}_S\left[\rho_S(\eta, t)X\right], \allowbreak \mathrm{Tr}_S\left[\rho_S(\eta, t) P\right])^T  $ and, $\bgreek{\sigma}(\eta ,t)$ is the covariance matrix (which is symmetric). Finally, $ \Psi(\eta, t) $ represents a time-dependent amplitude which is not conserved during the evolution due to the action of the non trace-preserving superoperator $\mathcal{L}_\eta(t)\left[\cdot\right]$.
Having assumed this ansatz for the characteristic function $\chi^{(\eta)}(q,p,t)$, Eq. \eqref{thetachar} can be equivalently expressed as
\begin{equation}\label{thetacharQBO}
\theta(t) = \frac{\partial \dot{\Psi}(\eta, t)}{\partial (i\eta)}|_{\eta=0}.
\end{equation}
Plugging Eq. \eqref{eq:ansatz} into Eq. \eqref{FokPl} and separating the different moments of q and p, it is easy to show that the evolution equation for the mean values $X_m(t), P_m(t)$ as well as for the off-diagonal elements of the covariance matrix $\sigma_{XP}(t)=\sigma_{PX}(t)$ has the following structure:
\begin{equation}
\partial_t O_i(t) = \sum_{j=1}^3 G_{ij} O_j(t),\quad\quad O_j \equiv \left\{X_m, P_m, \sigma_{XP}\right\},
\end{equation}
and therefore , provided we assume to deal with a system initially described by a thermal state so that $\sigma_{XP}(0) = X_m(0) = P_m(0) = 0$, we have that $\sigma_{XP}(t) = X_m(t) = P_m(t) = 0$ $\forall t$. Moreover $\sigma_{XX}(\eta, t) = \sigma_{PP}(\eta ,t) \equiv \sigma (\eta ,t)$ \cite{Puri,Olivares,Paternostro1}, and thus the number of evolution equations for the Gaussian parameters therefore reduce to the following two
\begin{align}\label{DiffResults}
&\partial_t \Psi(\eta,t) = \Psi(\eta,t) \left(2 V_1(\eta,t)\sigma(\eta, t) + V_2(\eta,t) \right)\\
&\partial_t \sigma(\eta,t) = \frac{1}{2}\left[2\Delta(t) + V_1(\eta,t)\right] \notag\\ 
&+ 2\left[V_2(\eta,t)-\gamma(t)\right]\sigma(\eta,t) + 2 V_1(\eta,t)\sigma^2(\eta,t)
\end{align}
We stress that, since $\lim_{\eta\to 0} V_j(\eta,t) = 0 \,\,(j=1,2)$, we retrieve in the case $\eta = 0$ the well-known solution for the characteristic function  \cite{Intravaia1,Intravaia2,Intravaia3,Adesso3}
\begin{equation}
\chi(q,p,t) = \exp\left[-\frac{q^2+p^2}{2} \sigma(0,t)\right]
\end{equation}
where $\sigma(0,t) \equiv \sigma(t)$ is given by Eq. \eqref{sigmasol} with the initial condition $\sigma(0) = 1/2\left(1+2 N(T_S)\right)$, with $N(T_S) = \left[\exp(1/T_S)-1\right]^{-1}$  ($T_S$ being the effective system's initial temperature).
From that, the final expression for the energy flow per unit of time given by Eq. \eqref{thetaFINAL} is easily obtained.

\section{Details on the numerical approach}
\label{app:B}

Here we give some details on the derivation of the evolution equation for the quadrature vectors $\mathbf{X}$ and $\mathbf{P}$ that lead, within the framework of a finite-size environment approach, to Eq. \eqref{ExNumEqs}. We stress that $X_{N+1}, P_{N+1}$ denote the position and momentum operators of the reduced system while the remaining $N$ operators refer to the environmental modes.
First of all, the couplings $g_i$ are determined through the definition of the spectral density \cite{Breuer2002} $J(\omega) = \sum_i \frac{|g_i|^2}{2\omega_i}\delta\left(\omega-\omega_i\right)$, which, by inversion, gives
\begin{equation}\label{couplings}
g_i = \pm\sqrt{2\omega_i \Delta\omega_i J(\omega_i)},\quad\quad\left(\Delta\omega_i \equiv \omega_i-\omega_{i-1}\right).
\end{equation}
Note that their sign is not uniquely determined by the spectral density but we arbitrarily take them to be positive.
We finally point out that the QBM studied in the previous Section is retrieved when we take the limit $N\to+\infty$, in which case however the numerical approach is not treatable.

As stated in the main body of the paper, since Eq. \eqref{HamNum} is quadratic in position and momentum, it can always be diagonalized by means of an orthogonal transformation $\mathbf{O}$ \cite{Vasile}, i.e. $\mathbf{M} = \mathbf{O}\mathbf{D}\mathbf{O}^T$ with $\mathbf{D}$ a diagonal matrix made of the eigenvalues $\lbrace \sqrt{2 d_i}\rbrace_{i=1,\ldots ,N+1}$ (often referred to as \textit{eigenfrequencies}) of $\mathbf{M}$.
By moving to the new coordinates $\tilde{\mathbf{X}} = \mathbf{O}^T \mathbf{X}$ and $\tilde{\vett{P}} = \mathbf{O}^T \mathbf{P}$, referred to as normal modes, we can express Eq. \eqref{HamNum} as
\begin{equation}
\Ham = \sum_{i=1}^{N+1} \frac{1}{2}\left(\tilde{P}_i^2 + d_i^2\tilde{X}_i^2\right),
\end{equation}
which leads to a free evolution
\begin{align}
&\tilde{X}_i (t) = \tilde{X}(0) \cos\left(d_i t\right) + \frac{\tilde{P}_i(0)}{d_i} \sin\left(d_i t\right) \\
&\tilde{P}_i (t) = - d_i \tilde{X}(0) \sin\left(d_i t\right) + \tilde{P}_i(0) \cos\left(d_i t\right).
\end{align}
Coming back to the original picture and defining the diagonal matrices $\mathbf{Cos}, \mathbf{Sin}$ and $\tilde{\mathbf{D}}$ we get the result in Eq. \eqref{ExNumEqs}.


\begin{thebibliography}{99}
\bibitem{footnote2} The results do not change with higher numbers of bath modes. Also, the Poincar\'{e} recurrence time is well beyond the considered propagation time.
\end{thebibliography}
\end{document}